\pdfoutput=1

\documentclass[prd,twocolumn,aps,amssymb,amsmath,tightenlines,floatfix]{revtex4}

\usepackage{graphicx,epsfig,xcolor}
\usepackage[T1]{fontenc}
\usepackage[latin1]{inputenc}
\usepackage[english]{babel}
\usepackage{amsmath}
\usepackage{mathtools}
\usepackage{amsfonts}
\usepackage{amssymb}
\usepackage{dsfont}
\usepackage{color}
\usepackage[labelformat=parens]{subfig}
\usepackage[justification=raggedright,font=footnotesize]{caption}

\usepackage[colorlinks = true, linkcolor = blue, urlcolor = blue,
citecolor = blue, anchorcolor = blue]{hyperref}

\newcommand{\cP}{\ensuremath{\mathcal{P}}}
\newcommand{\cL}{\ensuremath{\mathcal{L}}}
\newcommand{\cH}{\ensuremath{\mathcal{H}}}
\newcommand{\cD}{\ensuremath{\mathcal{D}}}
\newcommand{\cT}{\ensuremath{\mathcal{T}}}
\newcommand{\cC}{\ensuremath{\mathcal{C}}}
\newcommand{\cE}{\ensuremath{\mathcal{E}}}
\newcommand{\cPT}{\ensuremath{\mathcal{PT}}}
\newcommand{\cCPT}{\ensuremath{\mathcal{CPT}}}
\newcommand{\half}{\mbox{$\textstyle{\frac{1}{2}}$}}
\newcommand{\threehalf}{\mbox{$\textstyle{\frac{3}{2}}$}}
\newcommand{\fourth}{\mbox{$\textstyle{\frac{1}{4}}$}}

\newcommand{\vep}{\varepsilon}

\begin{document}
\title{Towards perturbative renormalization of $\phi^2(i\phi)^\vep$ quantum
field theory}
\author{Alexander Felski$^a$}\email{felski@thphys.uni-heidelberg.de}
\author{Carl M.~Bender$^b$}\email{cmb@wustl.edu}
\author{S.~P.~Klevansky$^a$}\email{spk@physik.uni-heidelberg.de}
\author{Sarben Sarkar$^c$}\email{sarben.sarkar@kcl.ac.uk}

\affiliation{
$^a$Institut f\"ur Theoretische Physik, Universit\"at Heidelberg, 69120
Heidelberg, Germany\\
$^b$Department of Physics, Washington University, St.~Louis, Missouri 63130,
USA\\
$^c$Department of Physics, King's College London, London WC2R 2LS, UK}

\begin{abstract}
In a previous paper it was shown how to calculate the ground-state energy
density $\cE$ and the $p$-point Green's functions $G_p(x_1,x_2,...,x_p)$ for the
$\cPT$-symmetric quantum field theory defined by the Hamiltonian density
$\cH=\half(\nabla\phi)^2+\half\phi^2(i\phi)^\vep$ in $D$-dimensional Euclidean
spacetime, where $\phi$ is a pseudoscalar field. In this earlier paper $\cE$ and
$G_p(x_1,x_2,...,x_p)$ were expressed as perturbation series in powers of $\vep$
and were calculated to first order in $\vep$. (The parameter $\vep$ is a measure
of the nonlinearity of the interaction rather than a coupling constant.) This
paper extends these perturbative calculations to the Euclidean Lagrangian $\cL=
\half(\nabla\phi)^2+\half\mu^2\phi^2+\half g\mu_0^2\phi^2\big(i\mu_0^{1-D/2}\phi
\big)^\vep-iv\phi$, which now includes renormalization counterterms that are
linear and quadratic in the field $\phi$. The parameter $g$ is a dimensionless
coupling strength and $\mu_0$ is a scaling factor having dimensions of mass.
Expressions are given for the one-, two, and three-point Green's functions, and
the renormalized mass, to higher-order in powers of $\vep$ in $D$ dimensions
($0\leq D\leq2$). Renormalization is performed perturbatively to second order in
$\vep$ and the structure of the Green's functions is analyzed in the limit $D\to
2$. A sum of the most divergent terms is performed to {\it all} orders in
$\vep$. Like the Cheng-Wu summation of leading logarithms in electrodynamics, it
is found here that leading logarithmic divergences combine to become mildly
algebraic in form. Future work that must be done to complete the perturbative
renormalization procedure is discussed.
\vskip 0.2cm
\today 
\end{abstract}
\maketitle

\section{Introduction}
\label{s1}
Since the publication of the first paper on $\cPT$ symmetry in 1998 \cite{r1},
in which the $\cPT$-symmetric quantum-mechanical Hamiltonian 
\begin{equation}
\label{E1}
H=p^2+x^2(ix)^\vep
\end{equation}
was introduced, this research area has become highly active. This model has been
studied in detail \cite{r1-1}, and much theoretical research has been done on
the mathematical structure of non-Hermitian quantum systems \cite{r1-2}.
Beautiful experiments have been performed in diverse areas of physics including
optics, photonics, lasers, mechanical and electrical analogs, graphene,
topological insulators, superconducting wires, atomic diffusion, NMR, fluid
dynamics, metamaterials, optomechanical systems, and wireless power transfer
\cite{r2,r3,r4,r5,r6,r7,r8,r9,r10,r11,r12,r13,r14,r14a}. 

This paper considers the generalization of (\ref{E1}) to quantum field theory in
$D$-dimensional Euclidean space. In an earlier paper \cite{r15} we examined the
corresponding field-theoretic Lagrangian density
\begin{equation}
\cL=\half(\nabla\phi)^2+\half\phi^2(i\phi)^\vep,
\label{E2}
\end{equation}
where $\phi$ is a (dimensionless) pseudoscalar field. Note that (\ref{E2}) is
manifestly $\cPT$-symmetric because $\phi\to-\phi$ under space reflection $\cP$
and $i\to-i$ under time reversal $\cT$. Just as (\ref{E1}) is a testbed of
$\cPT$-symmetric quantum mechanics, (\ref{E2}) is a natural model for the study
$D$-dimensional $\cPT$-symmetric bosonic field theories.

The Hamiltonian (\ref{E1}) launched the field of $\cPT$-symmetric quantum theory
because it has the surprising feature that if $\vep\geq0$, its eigenvalues are
all discrete, real, and positive even though it is not Dirac-Hermitian
\cite{r16,r17}. (A {\it Dirac-Hermitian} Hamiltonian obeys the symmetry
constraint $H=H^\dag$, where $\dag$ indicates combined complex conjugation and
matrix transposition.) Moreover, the quantum theory defined by (\ref{E1}) is
unitary (probability conserving) with respect to the adjoint $\cCPT$, where
$\cC$ is a linear operator satisfying the three simultaneous operator equations
\cite{r18}
\begin{equation}
\label{E3}
\cC^2=1,\quad[\cC,\cPT]=0,\quad[\cC,H]=0.
\end{equation}
Thus, while the constraint that a Hamiltonian be Dirac-Hermitian is sufficient
to define a consistent quantum theory, it is not necessary. In short,
$\cPT$-symmetric quantum theory is not in conflict with the axioms of
conventional Hermitian quantum theory; rather, it is a complex generalization
of Hermitian quantum theory.

Almost all of the theoretical research on $\cPT$ symmetry has focused on
$\cPT$-symmetric quantum mechanics and experimental studies of $\cPT$-symmetric
classical systems, but very few papers have focused on $\cPT$-symmetric quantum
field theory. We briefly review some of the earlier work on $\cPT$-symmetric
quantum field theory:

\begin{enumerate}
\item Early studies of $\cPT$-symmetric quantum field theory considered the
case $\vep=1$ in (\ref{E2}) \cite{r19,r20}. This $i\phi^3$ quantum field theory
had emerged in studies of Reggeon field theory \cite{r20a} and the Lee-Yang
edge singularity \cite{r20b}. Conventional diagrammatic perturbation theory
can be used to treat this cubic interaction: One introduces a coupling constant
$g$ in the interaction term $ig\phi^3$ and expands the physical quantities in
powers of $g$. However, one cannot use conventional perturbation theory for
other values of $\vep>0$, integer or noninteger.

\item $\cPT$-symmetric electrodynamics also has a cubic interaction \cite{r21}.
The Johnson-Baker-Willey (JBW) program for constructing a finite massless
electrodynamics fails because the zero of the beta function yields at best a
negative and perhaps a complex value of $\alpha$ because conventional Hermitian
quantum electrodynamics (QED) is not asymptotically free. However, the JBW
procedure works for the $\cPT$-symmetric version of QED because this theory is
asymptotically free and one obtains a reasonable positive numerical value for
$\alpha$.

\item Renormalizing a Hermitian quantum field theory often causes the
Hamiltonian of the theory to become non-Hermitian. This problem was observed in
the Lee model \cite{r22}, which is again a theory with a cubic interaction.
Pauli and K\"all\'en showed that upon renormalization, ghost states (states of
negative norm) arise, and appear to violate unitarity in scattering processes.
This problem remained unresolved until 2005, when it was shown that if one uses
the appropriate $\cPT$-symmetric inner product for the renormalized Lee-model
Hamiltonian, there are no ghost states and the unitarity of the theory becomes
manifest \cite{r23}.

\item Renormalizing the Hamiltonian for the Standard Model of particle physics 
induces what appears to be instability in the vacuum state. This is due to the
contribution of the top-quark loop integral \cite{r24}. Once again, the
renormalized Hamiltonian appears to be non-Hermitian. However, by using
$\cPT$-symmetric techniques it was shown by using a simple model-field-theory
argument that the vacuum state and the next few higher-energy states are
actually stable (have real energy) \cite{r25}.

\item Introducing higher-order derivatives in a quantum field theory in order to
make Feynman integrals converge also makes the Hamiltonian appear to be 
non-Hermitian. (Higher-order derivatives induce Pauli-Villars ghosts.) However,
$\cPT$-symmetric techniques resolve this problem. The simplest field-theory
model that exhibits this problem is the Pais-Uhlenbeck model and
$\cPT$-symmetric techniques demonstrate that this theory has no ghosts
\cite{r26}.

\item The double-scaling limit in quantum field theory [a correlated limit in
which the number $N$ of species in an ${\rm O}(N)$-symmetric field theory
approaches infinity as the coupling constant approaches a critical value]
appears to lead to an unstable field theory. For a quartic scalar field theory
the value of $g_{\rm crit}$ is negative and one might think that the resulting
$-\phi^4$ theory is unstable. However, the techniques of $\cPT$-symmetric
quantum theory demonstrate that such a theory is actually stable and has a real
positive spectrum \cite{r27,r28}. It was observed by K.~Symanzik that a
$-\phi^4$ theory is asymptotically free even though it does not have a local
gauge symmetry but he called this theory ``precarious" because it appears to be
unstable.

\item Time-like Liouville field theories appear to be unstable and 
non-Hermitian but the techniques of $\cPT$-symmetric quantum field theory can 
be used to argue that the Hamiltonians of such theories have real spectra and 
induce unitary time evolution \cite{r29}.

\item Studies of the $\cPT$-symmetric Dirac equation suggest that one may have
species oscillations and still have massless neutrinos \cite{r30,r31}.
Interestingly, recent measurements in Germany have halved the upper bound on
the mass of the electron neutrino \cite{r31a}.
\end{enumerate}

In the studies above some striking results were obtained but many of these
field-theory papers considered only simple zero-dimensional toy models and
one-dimensional quantum-mechanical analogs that suggest the possible behaviors
of $\cPT$-symmetric field theories. Thus, there is strong motivation for
developing general methods for solving quantum field theories defined by
non-Hermitian $\cPT$-symmetric Hamiltonians.

Until recently, $\cPT$-symmetric quantum field theory has remained beyond the
reach of comprehensive analytical study because of three technical problems that
had to be overcome when trying to solve the $\cPT$-symmetric quantum field
theory in (\ref{E2}):

\begin{enumerate}
\item Feynman perturbation theory works for the cubic case $\vep=1$ in
(\ref{E2}) but for other integer values of $\vep$ the Feynman diagrams must be
supplemented by nonperturbative contributions, which are nontrivial. Moreover,
when $\vep$ is noninteger, there are no Feynman rules at all so one cannot
perform a conventional coupling-constant expansion.

\item The functional integral for the partition function $Z=\int\cD\phi\,\exp
\big(-\int d^D x\,\cL\big)$, does not converge if $\vep>1$ unless the path of
integration in function space lies in appropriate infinite-dimensional Stokes
sectors in complex field space. Multidimensional Stokes sectors ({\it Lefshetz
thimbles} \cite{r31b}) are unwieldy.

\item In $\cPT$-symmetric quantum theory one must use the $\cC$ operator,
obtained by solving (\ref{E3}), in order to obtain matrix elements. It is
difficult to calculate $\cC$, so the prospect of calculating Green's functions
appears to be rather dim.
\end{enumerate}

In Ref.~\cite{r15} it is shown how to overcome these three problems by extending
the methods developed in Refs.~\cite{r32,r33,r34} for Hermitian quantum field
theories to non-Hermitian $\cPT$-symmetric quantum field theories. In brief,
instead of using a coupling-constant expansion, a perturbation expansion in
powers of the parameter $\vep$, which is a measure of the nonlinearity of the
theory, is performed and summation methods are then used to evaluate the $\vep$
series. For small $\vep$ the functional integral converges on the real axis in
complex-field space, and thus multidimensional Stokes sectors are not required
for convergence. This solves problems 1 and 2 above. Expanding in powers of
$\vep$ introduces complex logarithms in the functional integrand and $\cPT$
symmetry is then enforced by defining the complex logarithm properly:
\begin{equation}
\label{E4}
\log(i\phi)\equiv\half i\pi|\phi|/\phi+\half\log\big(\phi^2\big).
\end{equation}
The logarithms are now real, and the techniques for handling these logarithms 
to any order in powers of $\vep$ are based on methods that were introduced in
Refs.~\cite{r32,r33,r34}.

The apparent problem with the $\cC$ operator is actually not a problem if we are
calculating Green's functions. This is because in a theory with an unbroken
$\cPT$ symmetry the vacuum state is an eigenvalue of $\cC$ with eigenvalue 1:
$\cC|0\rangle=|0\rangle$. Since the Green's functions are vacuum expectation
values, we may ignore problem 3 entirely. This observation was first made in
Ref.~\cite{r36}.

Why does ${\cal L}$ in (\ref{E2}) define an interesting theory? Let us first
look at the cubic case. For a conventional $g\phi^3$ theory the ground-state
energy density $\cE$ is a sum of 3-vertex vacuum-bubble Feynman diagrams, and
the Feynman perturbation expansion has the form 
$$\cE=\sum A_n g^{2n}.$$
This series diverges and the coefficients $A_n$ all have the same sign. Thus, if
we perform Borel summation, we find a cut in the Borel plane, which implies that
$\cE$ is complex. Thus, the vacuum state is unstable. However, to obtain the
$\cPT$-symmetric cubic theory we replace $g$ by $ig$. Now, the perturbation
expansion {\it alternates} in sign and the Borel sum of the series is {\it
real}, the vacuum is stable, and the spectrum of the theory is bounded below.

The case of the quartic $\cPT$-symmetric theory is more elaborate. The
ground-state energy density for a conventional Hermitian quartic $g\phi^4$
quantum field theory has a Feynman perturbation expansion of the form
$$\cE=\sum B_n(-g)^n.$$
Again, this series is divergent, and since the perturbation coefficients $B_n$
all have the same sign, the series is alternating and Borel summable. The Borel
sum of the perturbation series yields a {\it real} value for the vacuum energy
density. This implies that the conventional quartic theory has a stable ground
state, as one would expect.

It may seem that the $\cPT$-symmetric quartic theory obtained by replacing $g$
with $-g$ is problematic because the perturbation series no longer alternates in
sign: If we Borel-sum the series, we find a cut in the Borel plane, which
suggests that $\cE$ is complex and that the vacuum state is unstable, as one
might intuitively expect with an upside-down potential. This conclusion is
false!

If one examines the functional integral for the partition function of the
theory, one sees that the perturbative contribution to $\cE$ (the Feynman
diagrams) must be supplemented by imaginary {\it nonperturbative} contributions
arising from two saddle points in complex function space. These additional
pure-imaginary contributions exactly cancel the discontinuity in the Borel plane
\cite{r35}. Consequently, the vacuum state for a $\cPT$-symmetric $-g\phi^4$
theory is stable. We emphasize that Feynman diagrams alone are not sufficient to
calculate the Green's functions of a $\cPT$-symmetric quantum field theory.

The research objectives in this paper are to extend the work in Ref.~\cite{r15}
and to study in depth the problem of renormalization. In Ref.~\cite{r15} the
Lagrangian (\ref{E2}) was examined to first order in $\vep$. Treating $\vep$ as
a small perturbation parameter, $\cL$ was expanded in a series, which to first
order in $\vep$ is
\begin{equation}
\cL=\half(\nabla\phi)^2+\half\phi^2+\half\vep\phi^2\log(i\phi)+{\rm O}\big(
\vep^2\big).
\label{E5}
\end{equation}
Identifying the free Lagrangian as 
\begin{equation}
\cL_0=\half(\nabla\phi)^2+\half\phi^2,
\label{E6}
\end{equation}
we developed techniques to evaluate the shift in the ground-state energy density
$\Delta E$ and the Green's functions $G_p$ to first order in $\vep$. We found
that
\begin{eqnarray}
\Delta E&=&\fourth\vep(4\pi)^{-D/2}\Gamma\big(1-\half D\big)
\nonumber\\
&&\times\,\big\{\log\big[2(4\pi)^{-D/2}\Gamma\big(1-\half D\big)\big]+\psi\big(
\threehalf\big)\big\},
\label{E7}\\
G_1&=&-i\vep\sqrt{\half\pi(4\pi)^{-D/2}\Gamma\big(1-\half D\big)}.
\label{E8}
\end{eqnarray}
We then found that the two-point connected Green's function in momentum space to
first order in $\vep$ is
$$\widehat{G}_2(p)=1/\big[p^2+1+\vep K+{\rm O}\big(\vep^2\big)\big],$$
where $K=\threehalf-\half\gamma+\half\log\big[\half(4\pi)^{-D/2}\Gamma\big(1-
\half D\big)\big]$. Thus, the renormalized mass to order $\vep$ is
\begin{equation}
\label{E9}
M_R^2=1+K\vep+{\rm O}\big(\vep^2\big).
\end{equation}

In addition, the higher-order connected Green's functions were also calculated
to first order in $\vep$:
\begin{eqnarray}
&&G_p(y_1,...,y_p)=-\half\vep(-i)^p\Gamma(\tfrac p2-1)\big[\half(4\pi)^{-D/2}
\nonumber\\
&&\!\!\!\!\!\!\!\!\!\times\,\Gamma\big(1-\half D\big)\big]^{1-p/2}\int d^Dx
\prod_{k=1}^p\Delta_1(y_k-x),
\label{E10}
\end{eqnarray}
where the free propagator $\Delta_\lambda(x)$ associated with $\cL_0=\half(
\nabla\phi)^2+\half\lambda^2\phi^2$ obeys the general $D$-dimensional Euclidean
Klein-Gordon equation 
\begin{equation}
\big(-\nabla^2+\lambda^2\big)\,\Delta_\lambda(x)=\delta^{(D)}(x).
\label{E11}
\end{equation} 
The solution to (\ref{E11}),
$$\Delta_\lambda(x)=\lambda^{D/2-1}|x|^{1-D/2}(2\pi)^{-D/2}K_{1-D/2}
(\lambda|x|),$$
has the property that
\begin{equation}
\label{E12}
\textstyle{\int}d^D x\,\Delta_\lambda(x)=\lambda^{-2}.
\end{equation}
The corresponding selfloop is then
\begin{equation}
\Delta_\lambda(0)=\lambda^{D-2}(4\pi)^{-D/2}\Gamma\big(1-\half D\big).
\label{E13}
\end{equation}
Selfloop factors with $\lambda=1$, which is associated with $\cL_0$ in
(\ref{E6}), enter into (\ref{E7}), (\ref{E8}), and (\ref{E10}) and lead to the
factors of $\Gamma\big(1-\half D\big)$ that occur in these expressions. This
calculation is verified in $D=0$ and $D=1$ in Ref.~\cite{r15}, where exact
calculations are possible, and provides confidence in these perturbative
results.
 
To proceed with the renormalization program we must overcome two problems.
First, we must calculate the Green's functions to higher order in $\vep$.
Second, we must show how to renormalize these Green's functions perturbatively
for $D\geq2$. This paper is focused on renormalization in two dimensions.
Specifically, we begin with the formulas for the one-point Green's function
$G_1$ in (\ref{E8}) and for the square of the renormalized mass $M_R^2$ in
(\ref{E9}). These quantities are finite for $0\leq D<2$. However, as $D$
approaches $2$ from below ($D\to2^-$) they diverge because $\Gamma(z)$ has a
pole at $z=0$: $\Gamma(z)\sim\frac1z$ as $z\to0$. Hence, $\Delta_1(0)$ in
(\ref{E13}) becomes infinite. To study the behavior of $G_1$ and $M_R^2$ near
$D=2$ we define $\delta\equiv2-D$; near $D=2$,
\begin{equation}
\label{E14}
\Delta_1(0)\sim\tfrac{1}{2\pi\delta}\quad(\delta\to0)
\end{equation} 
and the formulas for $G_1(\vep)$ and $M_R^2$ simplify to
\begin{eqnarray}
G_1(\vep)&\sim&-i\vep\tfrac1{2\sqrt\delta}\quad(\delta\to0),\label{E15}\\
M_R^2&\sim&-\half\vep\log\delta+A\quad(\delta\to0),\label{E16}
\end{eqnarray}
where $A=1+\vep\big[\threehalf-\frac{\gamma}{2}-\half\log(4\pi)\big]$. From
(\ref{E10}) we see that the Green's functions $G_p$ ($p>2$) vanish as
$\delta\to0$, so the theory becomes noninteracting to order $\vep$ at $D=2$. 

The question is whether perturbative renormalization can be accomplished in the
context of an expansion in powers of the parameter $\vep$. Ordinarily, for
interacting bosonic field-theories, renormalization is performed in the context
of a {\it coupling-constant} expansion. Here, we perform expansions in powers of
$\vep$, which is a measure of the nonlinearity of the selfinteraction. Series
expansions of this type were introduced many years ago \cite{r32} (prior to the
study of $\cPT$ symmetry), but the problem of renormalization has not been
addressed until now.

In our renormalization program, we use the information gained from (\ref{E15})
and (\ref{E16}). The one-point Green's function $G_1$, which is not directly
measurable, becomes infinite as $\delta\to0$. We can remove this divergence by
introducing in the Lagrangian a linear counterterm $iv\phi$, where $v$ has
dimensions of $({\rm mass})^{1+D/2}$ and $v=v_1\vep+v_2\vep^2+v_3\vep^3+\cdots$.
Such a term is consistent with $\cPT$ symmetry if $v$ is real: Under $\cPT$
reflection both the pseudoscalar field $\phi$ and $i$ change sign. In addition,
the divergence in (\ref{E16}) suggests that we should introduce an (infinite)
mass counterterm $\mu$ (the unrenormalized mass) into the Lagrangian.
Perturbative mass renormalization then consists of expressing the renormalized
mass $M_R$ in terms of these Lagrangian parameters and absorbing into the
parameter $\mu$ the divergence that arises as $\delta\to0$.

Thus, we consider the Lagrangian density 
\begin{equation}
\label{E17}
\cL=\half(\nabla\phi)^2+\half\mu^2\phi^2+\half g\mu_0^2\phi^2\big(i\mu_0^{1-D/2}
\phi\big)^\vep-iv\phi,
\end{equation}
that now contains a dimensional field $\phi$, the dimensional parameters $\mu$,
$v$, and $\mu_0$ (a fixed parameter having dimensions of mass), and the
dimensionless unrenormalized coupling $g$. Our objective is to calculate Green's
functions for this quantum field theory as series in powers of the parameter
$\vep$ and then to carry out perturbative renormalization for the
two-dimensional case. This is a nontrivial extension of the earlier work in
which the Green's functions were calculated to leading order in powers of $\vep$
\cite{r15} for the dimensionless Lagrangian density (\ref{E2}).

We thus generalize (\ref{E2})--(\ref{E16}) to first order in $\vep$. The formula
for $G_1$ in (\ref{E8}) is modified to read
\begin{equation}
\label{E18}
G_1(\vep)=-\tfrac{i\vep g}{m^2}\mu_0^{D/2-1}\sqrt{\half\pi m^{D-2}\Delta_1(0)}
+\tfrac{i\vep v_1}{\mu_0^2m^2}
\end{equation}
and the renormalized mass (\ref{E9}) becomes
\begin{equation}
\label{E19}
M_R^2=(m\mu_0)^2+\half\vep g\mu_0^2\big\{3-\gamma+\log\big[\half m^{D-2}
\Delta_1(0)\big]\big\},
\end{equation}
where in both expressions we introduce the dimensionless quantity $m^2=g+\mu^2/
\mu_0^2$ and we display the new parameters explicitly in terms of $\Delta_1(0)$,
whose behavior as $\delta\to0$ is given by (\ref{E14}). 

The formulas for $G_1(\vep)$ and $M_R^2$ simplify to
\begin{eqnarray}
G_1(\vep)&\sim&\tfrac{i\vep}{g\mu_0^2+\mu^2}\Big(v_1-\tfrac{g\mu_0^2}{2
\sqrt{\delta}}\Big)\quad(\delta\to0),
\label{E20}\\
M_R^2&\sim&\mu^2-\half\vep g\mu_0^2\log\delta+A\quad(\delta\to0),
\label{E21}
\end{eqnarray}
where $A=g\mu_0^2\big\{1+\vep\big[\threehalf-\textstyle{\frac{\gamma}{2}}-\half
\log(4\pi)\big]\big\}$ is a finite quantity having dimensions of $({\rm mass}
)^2$. By setting $v_1=g\mu_0^2/\big(2\sqrt{\delta})$, we remove the divergence
in $G_1$ as $\delta\to0$. Next, we see from (\ref{E21}) that the renormalized
mass $M_R$ is {\it logarithmically} divergent as $D\to2^-$. We absorb this
divergence into the mass counterterm $\mu$ by setting
\begin{equation}
\label{E22}
\mu^2=B+\half\vep g\mu_0^2\log\delta,
\end{equation}
where $B$ is a constant having dimensions of $({\rm mass})^2$. (Note that the
$\mu^2$ counterterm is large and {\it negative}.) The renormalized mass is now
finite, $M_R^2=A+B$, and the constant $B$ is in principle determined from the
experimental value of the renormalized mass $M_R$.

In this paper we study the expansions of the Green's functions to higher-order
in $\vep$ and examine perturbative renormalization in the limit $\delta\to0$. In
Sec.~\ref{s2} we use techniques developed in \cite{r15} to calculate the
connected Green's functions to order $\vep^2$. We examine the effects of the
infinite linear and quadratic counterterms $iv\phi$ and $\mu^2\phi^2$ on the
Green's functions to second order in $\vep$. In Sec.~\ref{s3} we perform a {\it
multiple-scale analysis} in which we obtain the leading contribution to each
order in $\vep$ and sum these terms to all orders in $\vep$. We compare the
results as $\delta\to0$ with those obtained via perturbative renormalization.
Conclusions are given in Sec.~\ref{s4}.

\section{Green's functions to order $\vep^2$}
\label{s2}
The $p$-point Green's function $G_p$ is defined as
\begin{equation}
G_p(\vep;y_1,...,y_p)=\frac 1Z \int\cD\phi\,e^{-\int d^D x\,\cL}\phi(y_1)...
\phi(y_p),
\label{E23}
\end{equation}
where the Lagrangian (\ref{E17}) enters both in the exponential function in the 
numerator and in the partition function in the denominator. Expanding the 
interaction term in powers of $\vep$, we get
\begin{eqnarray}
&&G_p(\vep;y_1, ...,y_p)=\frac1{Z_0}\int\!\cD\phi\,e^{-\int d^D x\,\cL_0} 
\phi(y_1)...\phi(y_p)\nonumber\\
&&\!\!\!\!\!\!\!\!\!\!\!\times\exp\bigg[\!-\frac{g\mu_0^2}{2}\sum_{n=1}^\infty
\frac{\vep^n}{n!}\!\int\!\!d^Dx\,\phi^2(x)\log^n\big(i\mu_0^{1-D/2}\phi\big)
\bigg],
\label{E24}
\end{eqnarray}
where
\begin{equation}
\cL_0=\half(\nabla\phi)^2+\half\mu_0^2m^2\phi^2.
\label{E25}
\end{equation}

The Green's function $G_p$ is not connected and to solve the renormalization
problem we require the {\it connected} $p$-point Green's functions, which are
constructed from cumulants. The procedure to order $\vep$ is explained in detail
in Ref.~\cite{r15} but to second order the cumulants become quite complicated.
Of course, the appropriate connected graphs are easy to identify if the
expression for $G_p$ contains polynomials and not logarithms of the field
$\phi$, and in this case standard techniques can then be used to evaluate the
graphs. Our strategy here is to recast (\ref{E24}) into a form containing only
products of the field $\phi$ at different space-time points. [The expression
(\ref{E24}) does not include higher-order terms in $\vep$ that arise from
expanding the denominator $Z$ in powers of $\vep$, as these lead to disconnected
diagrams.] From here on, in our Green's function calculations we always discard
disconnected contributions to the Green's functions, and we use the notation
$G_p$ to represent the {\it connected} Green's functions.

Let us consider terms in $G_p(\vep;y_1,...y_p)$ to order $\vep^2$:
\begin{eqnarray}
G_p(\vep;y_1,...,y_p) &=& \frac 1{Z_0}\int\cD\phi\,e^{-\int d^D x\,\cL_0} 
\phi(y_1)...\phi(y_p)\nonumber\\
&&\hspace{-3.0cm}\times\big[1+\vep g\mu_0^2 I_1+\tfrac 12\vep^2 g
\mu_0^2\big(I_2+g\mu_0^2 I_1^2\big)+{\rm O}\big(\vep^3\big)\big], 
\label{E26}
\end{eqnarray}
where
\begin{equation}
I_n=-\half{\textstyle\int}d^Dx\,\phi^2(x)\log^n\big[i\mu_0^{1-D/2}\phi(x)
\big].
\label{E27}
\end{equation}
Thus, in the expansion of the connected Green's functions in powers of $\vep$,
$$G_p(\vep;y_1,...,y_p)=\sum_{n=0}^\infty\vep^n G_{p,n}(y_1, ..., y_p),$$
we identify 
\begin{eqnarray}
G_{p,0}\!&=&\!\frac 1{Z_0}\int\!\!\cD\phi\,e^{-\!\int d^Dx\,\cL_0}\phi(y_1)...
\phi(y_p),\nonumber\\
G_{p,1}\!&=&\!\frac{g\mu_0^2}{Z_0}\int\!\!\cD\phi\,e^{-\!\int d^D x\,\cL_0}
\phi(y_1)...\phi(y_p)I_1,\label{E28}\\
G_{p,2}\!&=&\!\frac{g\mu_0^2}{2Z_0}\int\!\!\cD\phi\,e^{-\!\int d^Dx\,\cL_0}
\phi(y_1)...\phi(y_p)\big(I_2+g\mu_0^2 I_1^2\big),\nonumber
\end{eqnarray}
and so on, where we have suppressed the arguments $y_1,...,y_p$ on the left 
sides.

\subsection{One-point Green's function to second order}
\label{s2a}
The one-point Green's function can be calculated by adopting the techniques
developed in Ref.~\cite{r15}. Terms odd under $\phi\to-\phi$ integrate to zero,
so $G_{1,0}=0$, and $G_{1,1}$, the term proportional to $\vep$, is a constant
independent of $y_1$. The solution, generalized from Ref.~\cite{r15}, is
\begin{equation}
G_{1,1}=-i gm^{-2}\mu_0^{D/2-1}\sqrt{\pi m^{D-2}\Delta_1(0)/2},
\label{E29}
\end{equation}
which is given in (\ref{E18}). In Ref.~\cite{r15} it is explained how to treat
the term $I_1$ in (\ref{E28}). This term contains a complex logarithm in $I_1=
-\half\int d^Dx\,\phi^2(x)\log\big[i\mu_0^{1-D/2}\phi(x)\big]$ given in
(\ref{E27}). The complex logarithm is converted to a real logarithm via
(\ref{E4}) and the real logarithm is then treated by applying the replica trick
\cite{R7}
\begin{equation}
\label{E30}
\log\big(\mu_0^{2-D}\phi^2\big)=\lim_{N\to0}\,\tfrac d{dN}\big(\mu_0^{1-D/2}
\phi\big)^{2N},
\end{equation}
where the mass constant $\mu_0$ keeps the equation dimensionally consistent.

A second insight in \cite{r15} concerns terms containing the structure $i|\phi|/
\phi$ in (\ref{E4}). Here, we use the integral identity $|\phi|/\phi=(2/\pi)
\int_0^\infty(dt/t)\sin(t\phi)$ and then expand $\sin(t\phi)$ as a Taylor
series in powers of $\phi$:
\begin{equation} 
\label{E31}
\frac{|\phi|}{\phi}=\frac2\pi \int_0^\infty dt\!\sum_{\omega=0}^\infty
\frac{(-1)^{\omega}t^{2\omega}}{(2\omega+1)!}\,\phi^{2\omega+1}.
\end{equation}
[Note that the integration variable $t$ has dimensions of $(\rm mass)^{1-D/2}$
so this equation is dimensionally consistent.] Thus, the leading term in the
$\vep$ expansion of $G_1(\vep)$ is proportional to $\vep$ and was determined to
be (\ref{E18}).

The coefficient of $\vep^2$ in the expansion of $G_1(\vep)$ in (\ref{E28})
is a sum of two functional integrals $G_{1,2}=A_1+A_2$:
\begin{eqnarray}
A_1 &=& -\frac{g\mu_0^2}{4Z_0}\int\cD\phi\,e^{-\int d^Dx\,\cL_0}\phi(y_1)
\nonumber\\
&&~~~\times\int d^Dx\,\phi^2(x)\log^2\big[i\mu_0^{1-D/2}\phi(x)\big],\nonumber\\
A_2 &=& \frac{g^2\mu_0^4}{8Z_0}\int\cD\phi\,e^{-\int d^Dx\,\cL_0}\phi(y_1)
\nonumber\\
&&~~~\times\Big\{\int d^Dx\,\phi^2(x)\log\big[i\mu_0^{1-D/2}\phi(x)\big]
\Big\}^2.
\label{E32}
\end{eqnarray}
To evaluate $A_1$ we insert (\ref{E4}) into (\ref{E32}) to obtain real
logarithms and then insert the factor of $\phi^2(x)$. Discarding terms odd in
$\phi$, we get
\begin{eqnarray}
A_1 &=& -\frac{i\pi}{8Z_0} g\mu_0^2\int d^Dx\int\cD\phi\,e^{-\int d^Dx\,\cL_0}
\nonumber\\
&&~~\times\,\,\phi(y_1)|\phi(x)|\phi(x)\log\big[\mu_0^{2-D}\phi^2(x)\big].
\nonumber
\end{eqnarray}

Next, we use (\ref{E30}) and (\ref{E31}) to replace $|\phi(x)|$ and the
logarithm by a sum over products of fields:
\begin{eqnarray}
&&A_1=-\frac{i}{4Z_0}g\mu_0^2\int\!d^Dx\int\cD\phi\,e^{-\int d^Dx\,\cL_0}
\phi(y_1)\!\int_0^\infty\!\!\!dt\nonumber\\
&&\!\!\!\!\times\,\sum_{\omega=0}^\infty
\frac{(-1)^{\omega}t^{2\omega}}{(2\omega+1)!}\,\phi^{2\omega+3}(x)
\lim_{N\to0}\frac{d}{dN}\big[\mu_0^{1-D/2}\phi(x) \big]^{2N}.\nonumber
\end{eqnarray}
We use graphical techniques to do the functional integral: $A_1$ contains
products of the fields $\phi(y_1)\phi^{2\omega+2N+3}(x)$ and represents a free
propagator connecting $y_1$ to $x$ in $(2\omega+2N+3)$ ways, multiplied by
products of selfloops from $x$ to $x$; there are $\omega+N+1$ selfloops.
The functional integral yields the value $\Delta(y_1-x)\Delta^{\omega+N+1}(0)
(2\omega+2N+3)!!$ [For brevity we suppress the mass subscript $\lambda=\mu_0
m$ in the free propagator $\Delta$ defined in (\ref{E11}).]

We can now perform the $D$-dimensional integral over $x$ and use (\ref{E12}):
$\int d^D x\,\Delta(y_1-x)=\mu_0^{-2}m^{-2}$. (Note that the translation
invariance of $G_1(\vep)$ holds to second order in $\vep$.) Combining these
results, we get
\begin{widetext}
$$A_1=-\frac i 4{gm^{-2}}\sum_{\omega=0}^\infty\int_0^\infty\!\!\!dt\,
\frac{(-t^2)^\omega}{(2\omega+1)!}\Delta^{\omega+1}(0)
\lim_{N\to0}\frac{d}{dN}\big[\mu_0^{2-D}\Delta(0)\big]^{N}(2\omega+2N+3)!!.$$
Taking the derivative with respect to $N$ and the limit $N\to0$, this simplifies
to
\begin{equation}
A_1=-\frac i 4{gm^{-2}}\sum_{\omega=0}^\infty\int_0^\infty\!\!\!dt\,
\frac{(-t^2)^\omega}{(2\omega+1)!}\Delta^{\omega+1}(0) (2\omega+3)!!
\big\{\log \big[2\mu_0^{2-D}\Delta(0)\big]+\psi\big(\omega+\tfrac 52\big)\big\},
\label{E33}
\end{equation}
where we use the duplication formula for the digamma function $\psi(2z)=\half
\psi(z)+\half\psi\big(z+\half\big)+\log 2$. To evaluate the sum and integral we
use \cite{r15}
\begin{equation}
\int_0^\infty\!\!\!dt\,\sum_{\omega=0}^\infty\frac{(-t^2)^\omega}{(2\omega+1)!}
\Delta^{\omega+1}(0)(2\omega+3)!!=\Delta(0)\int_0^\infty\!\!\!dt\,
\big[3-\Delta(0)t^2\big]e^{-\Delta(0)t^2/2}=\sqrt{2\pi\Delta(0)}.
\label{E34}
\end{equation}
\end{widetext}

To calculate the second term in (\ref{E33}) we insert the integral
representation \cite{R8}
$$\textstyle{\psi(a)=\int_0^\infty dz\Big(\frac{e^{-z}}z-\frac{e^{-az}}
{1-e^{-z}}\Big)},$$
perform the sum over $\omega$ and the integral over $t$, and then note that
the $z$ integral gives the factor $\psi(2)$. Thus,
\begin{eqnarray}
&&\int_0^\infty\!\!\!\!dt\sum_{\omega=0
}^\infty \!\!\frac{(-t^2)^\omega}{(2\omega+1)!}\Delta^{\omega+1}(0)
(2\omega+3)!! \psi\big(\omega +\tfrac52\big)\nonumber\\
&&~~~~=\sqrt{2\pi\Delta(0)}\,\psi(2).
\label{E35}
\end{eqnarray}
Combining (\ref{E34}) and (\ref{E35}), we obtain $A_1$ from (\ref{E33}):
\begin{equation}
A_1=-\tfrac i2 gm^{-2}\sqrt{\tfrac{\pi\Delta(0)}2}\Big\{\log\big[2\mu_0^{2-D}
\Delta(0)\big]+\psi(2)\Big\}.
\label{E36}
\end{equation}

\begin{figure*}
\centering
\null\hfill
\subfloat[]{\includegraphics[width=0.35\textwidth]{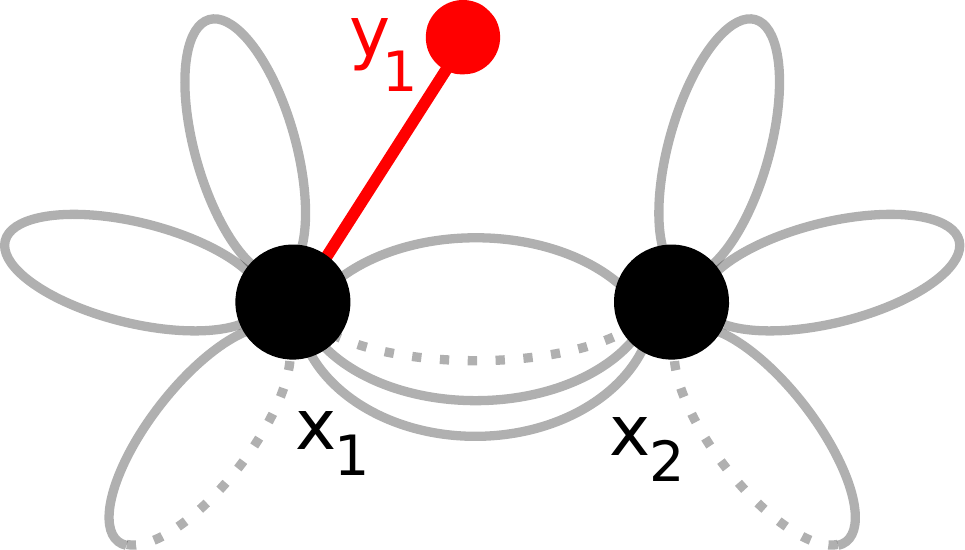}\label{F1a}}
\hfill
\subfloat[]{\includegraphics[width=0.35\textwidth]{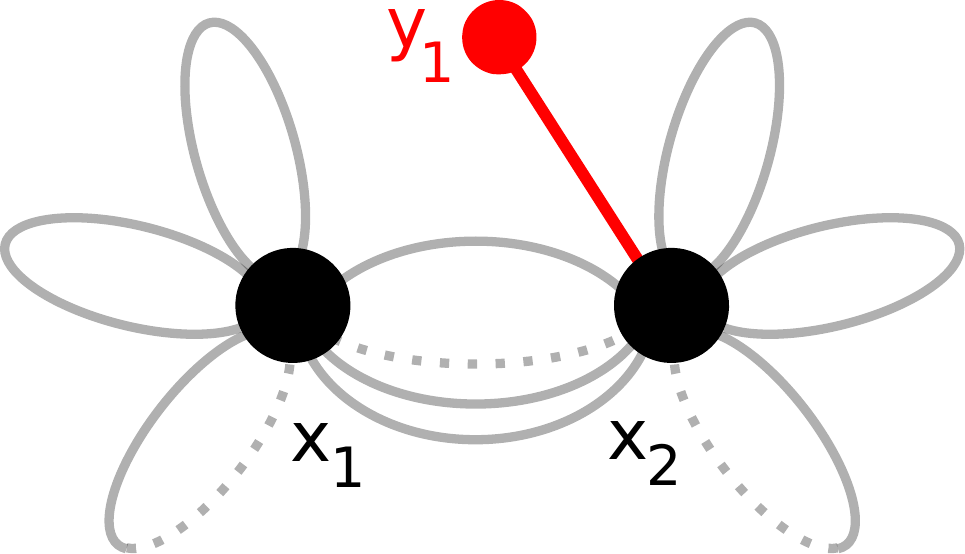}\label{F1b}}
\hfill\null
\caption{Graphical representation of terms contributing to the functional
integrals in (\ref{E37}). In \ref{F1a}, the line joining the external point
$y_1$ to the internal point $x_1$ is one of an {\it odd} number of lines meeting
at $x_1$; an even number of lines are connected to the internal point $x_2$. In
\ref{F1b} the line from the external point $y_1$ is one of an {\it even} number
of points at $x_2$; and {\it odd} number of lines connect to $x_1$. We integrate
over the internal points $x_1$ and $x_2$.} 
\end{figure*} 

Next, we evaluate $A_2$ in (\ref{E32}). Expanding $I_1^2$ produces fields $\phi$
at {\it two} points, say $x_1$ and $x_2$, which are integrated over. As before,
we replace each occurrence of an imaginary logarithm by using (\ref{E4}) and
retain terms that are even in $\phi$. (The functional integral vanishes for
terms that are odd in $\phi$.) Only one term survives:
\begin{eqnarray}
A_2 &=& \tfrac{i\pi g^2\mu_0^4}{16Z_0}\int\cD\,\phi e^{-\int d^D x\cL_0}
\phi(y_1)\int\! d^Dx_1\!\int\!d^D x_2\nonumber\\
&&~~\times\,\phi(x_1)|\phi(x_1)|\phi^2(x_2)
\log\big[\mu_0^{2-D}\phi^2(x_2)\big].\nonumber
\end{eqnarray}

Using (\ref{E31}) to replace $\phi(x_1)|\phi(x_1)|$ and the replica trick
(\ref{E30}) to replace the logarithm, we express $A_2$ as:
\begin{eqnarray}
A_2 &=& \frac{ig^2\mu_0^4}{8Z_0}\int\! d^Dx_1\!\int\!d^D x_2\int_0^\infty
\!\!\!dt\!\sum_{\omega=0}^\infty\frac{(-t^2)^\omega}{(2\omega+1)!}\nonumber\\
&&\!\!\!\!\!\!\!\!\!\!\!\!\times\,\int\!\cD\phi\,e^{-\int d^Dx\,\cL_0}\phi(y_1)
\phi^{2\omega+3}(x_1)\phi^2(x_2)\nonumber\\
&&\!\!\!\!\!\!\!\!\!\!\!\!\times\lim_{N\to0}\frac{d}{dN}\big[\mu_0^{1-D/2}
\phi(x_2)\big]^{2N}.
\label{E37}
\end{eqnarray}
Here we must evaluate the functional integral that connects the field at the
external point $y_1$ to one of the (odd number of) fields at the internal points
$x_1$; the remaining (even number of) points must be connected to the even
number of points at $x_2$. To this we must add the result of connecting the
external point $y_1$ to the (even number of) points at $x_2$ leaving an odd
number of connections from $x_2$ to $x_1$. This includes all connected diagrams
illustrated schematically in Figs.~\ref{F1a} and \ref{F1b}. 

The path integral corresponding to Fig.~\ref{F1a}, for the field combinations
$\phi(y_1)\phi^{2\omega+3}(x_1)\phi^{2N+2}(x_2)$ allows $y_1$ to connect to one
of the $2\omega+3$ replicas of $\phi$ at $x_1$. For the remaining points to be
connected, $2l$ lines can connect the remaining $2\omega+2$ points at $x_1$ to
the $2N+2$ points at $x_2$; the remainder of points $(2N+2-2l)+(2\omega+2-2l)$
are used to form $N+\omega+2-2l$ closed loops. Here, $l\ge1$ for the graph to be
connected. We assign to Fig.~\ref{F1a} the combinatoric factor 
$$C_a=\textstyle{\frac{(2N+2)!(2\omega+3)!}{(2l)!(N+1-l)!(\omega+1-l)!2^{N
+\omega+2-2l}}}.$$

For Fig.~\ref{F1b} there are $2N+2$ ways to connect $y_1$ to $x_2$. To create
connected graphs an odd number, say $2l+1$, of lines must join the remaining $2N
+1$ points at $x_1$ to the $2\omega+3$ points at $x_2$, where the minimum value
of $l$ is zero. Closed loops can exist on $2N+1-(2l+1)=2N-2l$ points on $x_2$
and $2\omega+3-(2l+1)$ points on $x_1$, forming a total of $N+\omega-2l+1$
loops. The combinatoric factor assigned to this graph is
$$C_b=\textstyle{\frac{(2N+2)!(2\omega+3)!}{(2l+1)!(N-l)!(\omega+1-l)!2^{N
+\omega+1-2l}}}.$$

\begin{widetext}

It is best to examine the contributions from Fig.~\ref{F1a} and Fig.~\ref{F1b}
separately. We use the combinatoric factor $C_a$ associated with Fig.~\ref{F1a}
to evaluate its contribution to (\ref{E37}) and perform one spatial integral:
\begin{eqnarray}
A_2^{\rm Fig.\,1a} &=& \frac{ig^2\mu_0^2}{8m^2}\lim_{N\to 0}\frac{d}{dN}
(2N+2)!\big[\half\mu_0^{2-D}\Delta(0)\big]^N\int d^Dx\int_0^\infty dt
\sum_{\omega=0}^\infty(-t^2)^\omega(2\omega+3)
(2\omega+2)\big[\half\Delta(0)\big]^{\omega+2}\nonumber\\
&&\times\sum_{l=1}^{min(N+1,\,\omega+1)}\frac{1}{(2l)!(N+1-l)!(\omega+1-l)!}
\Big[2\tfrac{\Delta(x)}{\Delta(0)}\Big]^{2l}.
\label{E38}
\end{eqnarray}
Exchanging the sums on $l$ and $\omega$ with $\omega\ge l-1$, we determine the
sum over $\omega$:
$$\sum_{\omega=l-1}^\infty(-z)^\omega \frac{(2\omega+3)(2\omega+2)}{(\omega
+1-l)!}=(-z)^{l-1}e^{-z}\big[4z^2-z(8l+6)+2l(2l+1)\big],$$
where we have shifted the summation variable to $p=\omega+1-l$, which runs from
$0$ to $\infty$, and we have used the sums $\sum_p z^p/p!=e^z$, $\sum_p pz^p/p!=
ze^z$, and $\sum_p p^2z^p/p!=(z^2+z)e^z$.

Setting $z=\half\Delta(0)t^2$ we evaluate the integral over $t$:
\begin{equation}
\int_0^\infty\!\!dt\sum_{\omega=l-1}^\infty\big[-\half\Delta(0)t^2\big]^\omega
\frac{(2\omega+3)(2\omega+2)}{(\omega+1-l)!}
=2\frac{(-1)^{l-1}}{\sqrt{2\Delta(0)}}\Gamma\big(l-\half\big).
\label{E39}
\end{equation}
Thus, (\ref{E38}) becomes
\begin{equation}
A_2^{\rm Fig.\,1a}=\frac{ig^2\mu_0^2}{4\sqrt2 m^2}[\Delta(0)]^{3/2}
\lim_{N\to 0}\frac{d}{dN}(2N+2)!\big[\half\mu_0^{2-D}\Delta(0)\big]^N
\int d^Dx\sum_{l=1}^\infty
\big[\tfrac{\Delta(x)}{\Delta(0)}\big]^{2l} 
\frac{(-1)^{l-1}\Gamma\big(l-\half\big)}{2^{2-2l}(2l)!(N+1-l)!}.
\label{E40}
\end{equation}

To evaluate the sum over $l$ we simplify the factorials and Gamma functions by
using the duplication formula $\Gamma(2z)=\Gamma(z)\Gamma\big(z+\half\big)2^{2z
-1}/\sqrt{\pi}$ and Euler's reflection formula $\Gamma(z)\Gamma(1-z)=\pi/\sin(
\pi z)$. Then (\ref{E40}) reduces to
\begin{equation}
A_2^{\rm Fig.\,1a}=\frac{ig^2\mu_0^2}{m^2}\big[\half\Delta(0)\big]^{3/2}
\lim_{N\to0}\frac{d}{dN}[2\mu_0^{2-D}\Delta(0)]^N\Gamma\big(N+\tfrac32\big)
\int d^Dx \Big\{{}_2F_1\big(-\half,-N-1;\half;\Big[\tfrac{\Delta(x)}{\Delta(0)}
\Big]^2\big)-1\Big\},
\label{E41}
\end{equation}
where ${}_2F_1\big(a,b;c;x^2\big)$ is a Gaussian hypergeometric function
\cite{R8}. When $N=0$ this function becomes simply ${}_2F_1\big(-
\half,-1;\half;x^2\big)=1+x^2$ and
$\lim_{N\to 0}\frac{d}{dN}\,{}_2F_1\big(-\half,-N-1;\half;x^2\big) 
=(1+x)^2\log(1+x)+(1-x)^2\log(1-x)-2x^2$.

Next, evaluate the contribution from Fig.~\ref{F1b}. Multiplying the
combinatorial factor $C_b$ with the associated propagators in the evaluation of
the functional integral of (\ref{E37}) and performing one spatial integration,
we get
\begin{eqnarray}
A_2^{\rm Fig.\,1b} &=& \frac{ig^2\mu_0^2}{8m^2}\lim_{N\to 0}\frac{d}{dN}
(2N+2)!\big[\half\mu_0^{2-D}\Delta(0)\big]^N\int d^Dx\int_0^\infty
dt\sum_{\omega=0}^\infty(-1)^\omega t^{2\omega}
(2\omega+3)(2\omega+2)\big[\half\Delta(0)\big]^\omega\nonumber\\
&&\times\sum_{l=0}^{min(N,\omega+1)}[\Delta(x)]^{2l+1}\big[\half\Delta(0)
\big]^{1-2l}\frac{1}{(2l+1)!(N-l)!(\omega+1-l)!}\nonumber
\end{eqnarray}
Exchanging orders of summation over $l$ and $\omega$ and using (\ref{E39}) to
sum over $\omega$ and integrate over $t$, we obtain
$$A_2^{\rm Fig.\,1b}=\frac{ig^2\mu_0^2}{4\sqrt 2m^2}[\Delta(0)]^{3/2}
\lim_{N\to 0}\frac{d}{dN}[\half\mu_0^{2-D}\Delta(0)]^{N}(2N+2)!\int d^Dx
\sum_{l=0}^\infty\Big[\tfrac{\Delta(x)}{\Delta(0)}\Big]^{2l+1} 
\frac{ (-1)^{l-1}\Gamma(l-\tfrac 12)}{2^{1-2l} (2l+1)!(N-l)!}.$$
Again, on applying the duplication and reflection formulas for the Gamma
function, we find that this expression also simplifies to the compact form
\begin{equation}
A_2^{\rm Fig.\,1b}=\frac{2ig^2\mu_0^2}{m^2}\big[\half\Delta(0)\big]^{3/2}
\lim_{N\to 0}\frac{d}{dN}[2\mu_0^{2-D}\Delta(0)]^N (N+1)\Gamma\big(N+\tfrac32
\big)\int d^Dx\frac{\Delta(x)}{\Delta(0)}\,
{}_2F_1\Big(-\half,-N;\tfrac32;\big[\tfrac{\Delta(x)}{\Delta(0)}\big]^2\Big).
\label{E42}
\end{equation}

When $N=0$, the hypergeometric function simplifies, ${}_2F_1\big(-\half,0;
\threehalf;x^2\big)=1$, and its derivative at $N=0$ is
$\lim_{N\to 0}\,\frac{d}{dN}{}_2F_1\big(-\half,-N;\,\threehalf;x^2\big)=
\frac{1}{2x}[(1+x)^2\,\log(1+x)-(1-x)^2\,\log(1-x)-2x].$
Thus, $A_2=A_2\,[{\rm Fig.\,1a}]+A_2\,[{\rm Fig.\,1b}]$ is constructed by
adding (\ref{E41}) and (\ref{E42}), yielding the formal result
\begin{eqnarray}
A_2&=&\frac{ig^2\mu_0^2}{m^2}[\tfrac12{\Delta(0)}]^{3/2}
\lim_{N\to0}\frac{d}{dN}[2\mu_0^{2-D}\Delta(0)]^{N}\,\Gamma\big(N
+\threehalf\big)\nonumber\\
&\times&\int d^D x\Big[\big ({}_2F_1\Big(-\half,-N-1;\half;\big[
\tfrac{\Delta(x)}{\Delta(0)}\big]^2\,\Big)-1\big)+2(N+1)
\tfrac{\Delta(x)}{\Delta(0)}\,{}_2F_1\Big(-\half,-N;
\threehalf;\Big[\tfrac{\Delta(x)}{\Delta(0)}\Big]^2\Big)\Big]\nonumber\\
&=&\tfrac{ig^2\mu_0^2\sqrt\pi}{2m^2}\big[\half\Delta(0)\big]^{3/2}
\Big\{\big[\log[2\mu_0^{2-D}\Delta(0)]+\psi\big(\threehalf\big)\big]
\!\!\int d^Dx\Big(\big[\tfrac{\Delta(x)}{\Delta(0)}\big]^2
+2\tfrac{\Delta(x)}{\Delta(0)}\Big) \nonumber\\
&& \quad+2\int d^Dx\,\Big[
\big(1+\tfrac{\Delta(x)}{\Delta(0)}\big)^2\log\big[1+\tfrac{\Delta(x)}
{\Delta(0)}i\big]-\big[\tfrac{\Delta(x)}{\Delta(0)}\big]^2\Big] \Big\}.\nonumber
\end{eqnarray}

The integrals containing linear and quadratic powers of $\Delta(x)$ are readily
evaluated, the linear one following from (\ref{E12}), and the quadratic one from
the solution to (\ref{E10}), using $\int_0^\infty dt\,tK^2_\nu(t)=\half\Gamma(
1-\nu)\Gamma(1+\nu)$. This yields $\int d^Dx\,[\Delta(x)]^2=(\mu_0 m)^{D-4}2^{
-D}\pi^{-D/2}\Gamma(2-D/2)$. Thus,
$$A_2=\tfrac i8 g^2m^{-4}\sqrt{\pi\Delta(0)/2}
\big\{\big[\log[2\mu_0^{2-D}\Delta(0)]+\psi(\tfrac32)\big](6-D)+2D-4+
4\Delta(0)\mu_0^2m^2\!\int\!d^Dx\,\big(1+\tfrac{\Delta(x)}{\Delta(0)}\big)^2
\log\big[1+\tfrac{\Delta(x)}{\Delta(0)}\big]\big\}.$$
The {\it connected} part of $G_1$ to order $\vep^2$ is $A_1+A_2$, so we add the
above result to (\ref{E36}) to get
\begin{eqnarray}
G_{1,2} &=& -\half igm^{-2}\sqrt{\half\pi\Delta(0)}\Big\{\log[2\mu_0^{2-D}
\Delta(0)]+\psi(2)\Big\}+\tfrac18 ig^2 m^{-4}\sqrt{\half\pi\Delta(0)}
\Big\{\big(\log[2\mu_0^{2-D}\Delta(0)]+\psi(\tfrac 32)\big)(6-D)\nonumber\\
&&\qquad+2D-4+4\Delta(0)\mu_0^2m^2\int d^Dx\,\big(1+\tfrac{\Delta(x)}
{\Delta(0)}\big)^2\log\big[1+\tfrac{\Delta(x)}{\Delta(0)}\big]\Big\}.\nonumber
\end{eqnarray}
\end{widetext}
The first term is proportional to the dimensionless coupling strength $gm^{-2}$
while the second term is proportional to its square. In the second term the
expression in curly brackets is a dimensionless number.

We now examine the limit $D\to2$ (that is, $\delta\to0$). From the solution to
(\ref{E11}) and (\ref{E13}) we see that the combination $\Delta(x)/\Delta(0)\sim
\delta K_0(\mu_0m|x|)$ to first order in $\delta$ as $\delta\to0$. Thus, to
lowest order in $\delta$ the last term containing the integral simplifies to
\begin{eqnarray}
&&4\Delta(0)\mu_0^2m^2\int d^Dx\, \Big(1+\frac{\Delta(x)}{\Delta(0)}
\Big)^2 \log\Big[1+\tfrac{\Delta(x)}{\Delta(0)}\Big]\nonumber\\
&&\sim 4\mu_0^2m^2 \int_0^\infty dx\,xK_0(\mu_0 m |x|)=4\nonumber
\end{eqnarray}
because $\int_0^\infty dt\,t^{\alpha-1}K_\nu(t)=2^{\alpha-2}\Gamma(\tfrac{\alpha
-\nu}{2})\Gamma(\frac{\alpha+\nu}{2})$, ${\rm Re}\,\alpha>|{\rm Re}\,\nu|$.
Thus, if $D\to2$, the coefficient of $G_1$ at second order in $\vep$ goes as 
\begin{eqnarray}
G_{1,2}&\sim& -\tfrac i4 gm^{-2}\delta^{-1/2}[\psi(2)-\log(\pi\delta)]
\nonumber\\
&&\!\!\!\!\!\!+\tfrac{i}{4}g^2m^{-4}\delta^{-1/2}\big[1+\psi(\tfrac 32)-
\log\pi-\log\delta\big]+{\rm O}(\delta).\nonumber
\end{eqnarray}

Taking the same limit in (\ref{E29}), the coefficient of $G_1$ to first order in
$\vep$ yields $G_{1,1}\to -\half igm^{-2}\delta^{-1/2}$. Putting these two last
results together, we find that the divergence structure of $G_1$ in the $\vep$
expansion has the form
$$ G_1\to-ic_1\delta^{-1/2}\vep -i\delta^{-1/2}(c_2+c_3\log\delta)\vep^2
+{\rm O}(\vep^3),$$
where $c_1=\half gm^{-2}$, $c_2=\fourth gm^{-2}\big[\psi(2)-\log\pi-gm^{-2}
\big(1+\psi\big(\tfrac32\big)-\log\pi\big)\big]$, and $c_3=-\fourth gm^{-2}(1-g
m^{-2})$ are constants. So, the algebraic divergence $\delta^{-1/2}$ occurs at
each order in the $\vep$ expansion. Our key result is that {\it the second-order
term in the $\vep$ expansion introduces $\log\delta$ but does not alter the
algebraic structure of the divergence.}

{\it Inclusion of the counterterm $-iv\phi$.} As seen from the calculations
above, $G_1$ is negative imaginary and diverges in two dimensions as $\delta\to
0$. Since $G_1$ is not a physically measurable quantity, it can be removed by
introducing an appropriate counterterm $-iv\phi$ into the Lagrangian, where
$v=v_1\vep+v_2\vep^2+{\rm O}(\vep^3)$.

To examine the effect of including such a term, we denote the one-point
connected Green's function associated with the full Lagrangian that includes
$-iv\phi$ as $G_1(v)$. Its relation to the Green's functions evaluated without
$v$ is determined as follows. [We keep the notation of this section, without
explicitly writing $v=0$; that is, $G_1(v=0)=G_1$ and similarly for the
higher-order Green functions and all expansion coefficients.] We include the
contributions to the path integral of $\exp\big[i\int d^Dx\big(v_1\vep+v_2\vep^2
\big)\phi(x)\big]$ to second order in $\vep$. This multiplies the expansion in
$\vep$ of $G_p$ in (\ref{E26}). Identifying terms proportional to $\vep$ and
$\vep^2$ and using the definitions in (\ref{E28}), we find that for any $p$
\begin{eqnarray}
G_p(v) &=& G_{p,0}+\vep\Big[G_{p,1}+iv_1\!\int\!\!d^Dx\,G_{p+1,0}(y_1,...y_p,x)
\Big]\nonumber\\
&+& \vep^2\Big[G_{p,2}+iv_2\int d^Dx\,G_{p+1,0}(y_1,...,y_p,x) 
\nonumber\\
&-& \frac 12 v_1^2\int d^Dx_1\int d^D x_2 G_{p+2,0} (y_1,...,y_p,x_1,x_2)
\nonumber\\
&+& iv_1 \int d^Dx G_{p+1,1}(y_1,...,y_p,x)\Big]+{\rm O}\big(\vep^3\big),
\label{E43}
\end{eqnarray}
where we have suppressed the arguments $y_1, ...,y_p$ of $G_{p,n}$. Thus, the
coefficients of the $\vep$ expansion $G_p(v)=\sum_n\vep^n G_{p,n}(v)$ are
related to the coefficients of the expansion of $G_p$ through the above
expression. In principle, if we calculate the coefficients in the $\vep$
expansion of the Green's function without $v$, we can easily determine the
effects of including it.

Let us evaluate $G_1(v)$. From (\ref{E43}) we get
\begin{eqnarray}
G_1(v) &=& \vep\Big[G_{1,1}+iv_1\int d^Dx\,G_{2,0}(y,x)\Big]+\vep^2\Big[G_{1,2}
\nonumber\\
&&\!\!\!\!\!\!\!\!\!\!\!\!\!\!\!+iv_2\int d^D x\,G_{2,0}(y,x)
+iv_1\int d^Dx\,G_{2,1}(y,x)\Big],\nonumber
\end{eqnarray}
since $G_{p,0}=0$ unless $p=2$. Thus, to evaluate $G_1(v)$ to second order in
$\vep$, we must know the expansion coefficients of the two-point Green's
function, calculated to first-order in $\vep$ without $v$. But these are known;
by definition, $G_{2,0}(y,x)=\Delta(y-x)$ while
\begin{equation}
G_{2,1}(y,x)=-g\mu_0^2K_1\int d^Dz\,\Delta(y-z)\Delta(z-x),
\label{E44}
\end{equation}
with $K_1=3/2-\gamma/2+\half\log\big[\mu_0^{2-D}\Delta(0)/2\big]$ is a
generalization of the result obtained in \cite{r15} that includes the mass
parameter $\mu_0 m$ of the Lagrangian $\cL_0$ used here [see (\ref{E25})]. In
the limit $\delta\to 0$, $G_{1,1}\to-igm^{-2}/(2\sqrt\delta)$ from (\ref{E29}),
so in this limit
\begin{eqnarray}
G_1(v) &\sim& i\vep\Big(\frac{v_1}{\mu_0^2m^2}-\frac{g}{2m^2\sqrt\delta}\Big)
\nonumber\\
&&\!\!\!\!\!\!\!\!\!\!\!\!\!\!\!\!\!\!\!\!\!\!\!\!\!\!\! +i\vep^2\Big[\frac{v_2}
{\mu_0^2m^2}-\frac{gK_1v_1}{\mu_0^2m^4}-\delta^{-1/2}(c_2+c_3\log\delta)\Big].
\label{E45}
\end{eqnarray}
Setting $G_1(v)=0$, we obtain the first-order result in (\ref{E20}), which fixes
$v_1\to g\mu_0^2/(2\sqrt\delta)$. If we insert this into (\ref{E45}), then $v_2$
in turn is fixed to eliminate the ${\rm O}\big(\vep^2\big)$ term. It too has the
same divergence structure $v_2\to\delta^{-1/2}$.

In summary, from our ${\rm O}\big(\vep^2\big)$ calculation of $G_1$, we find
that the divergence structure obtained when $D\to2$ has the {\it same} algebraic
form that was determined from the ${\rm O}(\vep)$ term; namely, $\delta^{-1/2}$.
It is accompanied by a logarithmic divergence in $\delta$. This is a structure
that persists for higher-order Green's functions. We will see that as $D\to2$,
the algebraic structure of the lowest-order Green's functions is accompanied by
logarithmic divergences as one goes to higher orders in the $\vep$ expansion. 

\subsection{Two-point Green's function in second order}
\label{s2b}
A general expression for the $p$-point Green's functions and their coefficients
in the $\vep$ expansion, given in (\ref{E26})--(\ref{E28}) can be formulated
using the generalized form of the replica trick (\ref{E30})
$$\log^m\big[\mu_0^{2-D}\phi^2\big]=\lim_{N\to0}\,(\tfrac d{dN})^m\big(\mu_0^{1
-D/2}\phi\big)^{2N}$$
and the generalization of (\ref{E31}),
$$\Big(\frac{|\phi|}{\phi}\Big)^m=\frac2\pi\int_0^\infty\!\!dt\sum_{\omega=0}^
\infty\frac{\big(-t^2\big)^\omega}{(2\omega+1)!}\,\phi^{(2\omega+1)m}.$$
Before applying these, expressions containing $I_n$ will have the occurrence of
$\log^n\big[i\mu_0^{1-D/2}\phi(x)\big]$ in (\ref{E27}) written in terms of
real logarithms via (\ref{E4}) and expanded in a binomial series. As illustrated
in detail in the previous section, the aim is to reduce the expressions that
arise to products of functional integrals containing powers of $\phi$ multiplied
by appropriate factors. 

\begin{figure}[t]
\vspace{.6cm}
\includegraphics[width=0.24\textwidth]{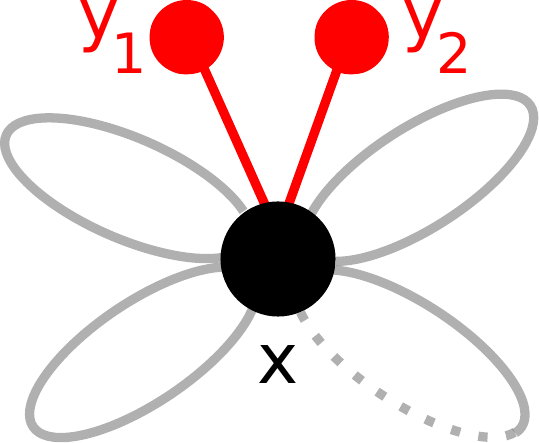}
\caption{Graphical representation of connected terms contributing to the
functional integrals in (\ref{e46}) and (\ref{E47}).} 
\label{F2}
\end{figure}

From (\ref{E28}) the lowest-order contribution to $G_2(y_1,y_2)$ is the free
Green's function $G_{2,0}(y_1,y_2)=\Delta(y_1-y_2)$. Performing the operations
delineated in the last section, the coefficient of $\vep$ becomes 
\begin{eqnarray}
G_{2,1}(y_1,y_2) &=& -\frac{g\mu_0^2}4\int d^Dx \lim_{N\to0}\frac d{dN}
\mu_0^{(2-D)N}\nonumber \\
&&\!\!\!\!\!\!\!\!\!\!\!\!\!\!\!\!\!\!\!\!\!\!\!\!\!\!\!\!\times\int
\frac{\cD\phi}{Z_0}\,e^{-\int d^Dx\,\cL_0}\phi(y_1) \phi(y_2)\phi^{2N+2}(x),
\label{e46}
\end{eqnarray}
where the connected graph associated with the functional integral connects the
two external points to one internal one (see Fig.~\ref{F2}). In this case the
functional integral takes the value $[(2N+2)!/(2^N N!)][\Delta(0)]^N\Delta
(y_1-x)\Delta(x-y_2)$, leading to the final result given in (\ref{E44}).

\begin{figure*}
\centering
\null\hfill
\subfloat[]{\includegraphics[width=0.22\textwidth]{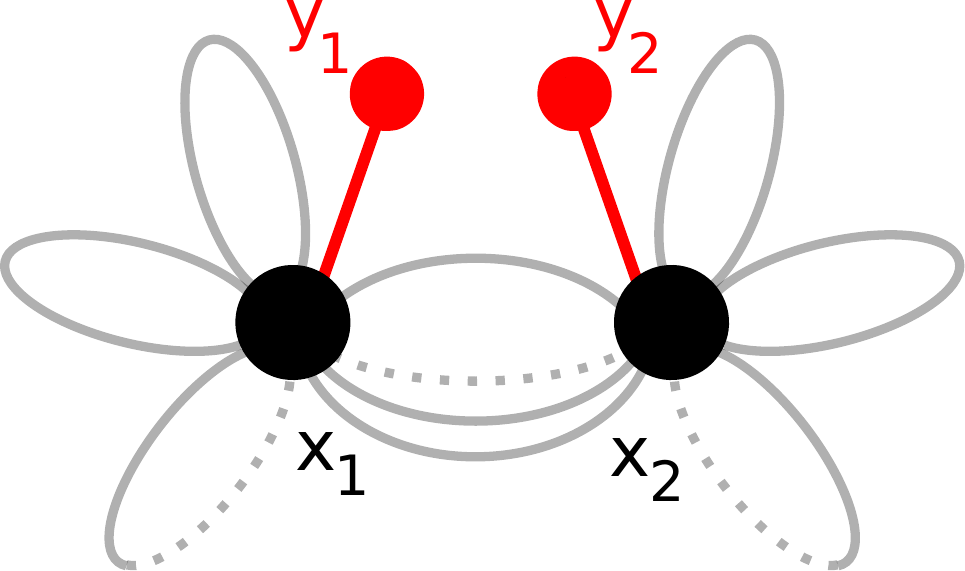}\label{F3a}}
\hfill
\subfloat[]{\includegraphics[width=0.22\textwidth]{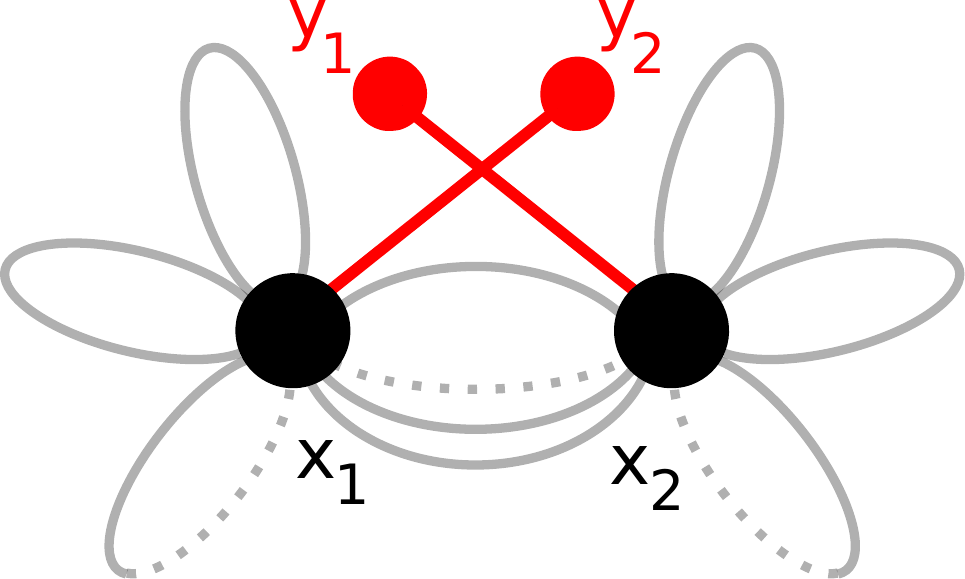}\label{F3b}}
\hfill
\subfloat[]{\includegraphics[width=0.22\textwidth]{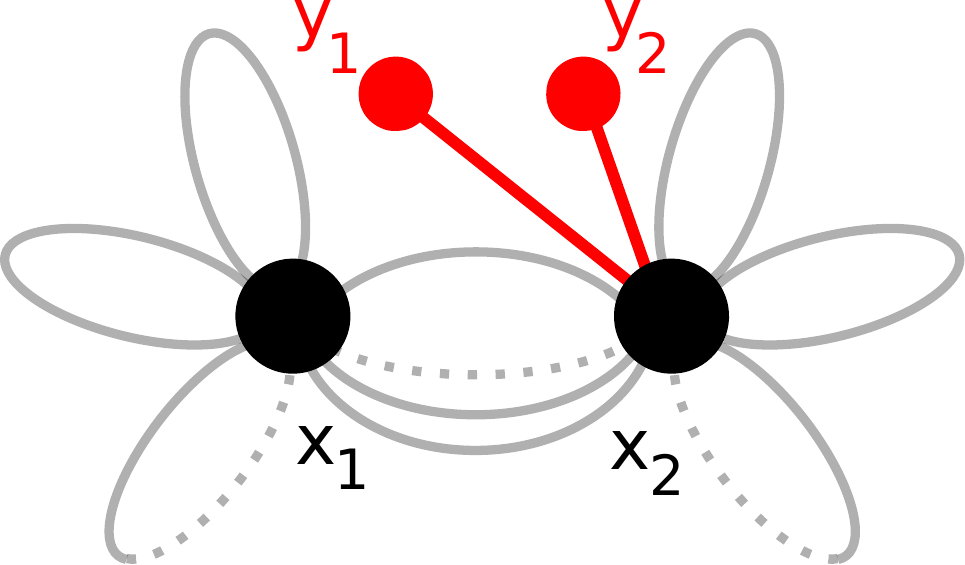}\label{F3c}}
\hfill
\subfloat[]{\includegraphics[width=0.22\textwidth]{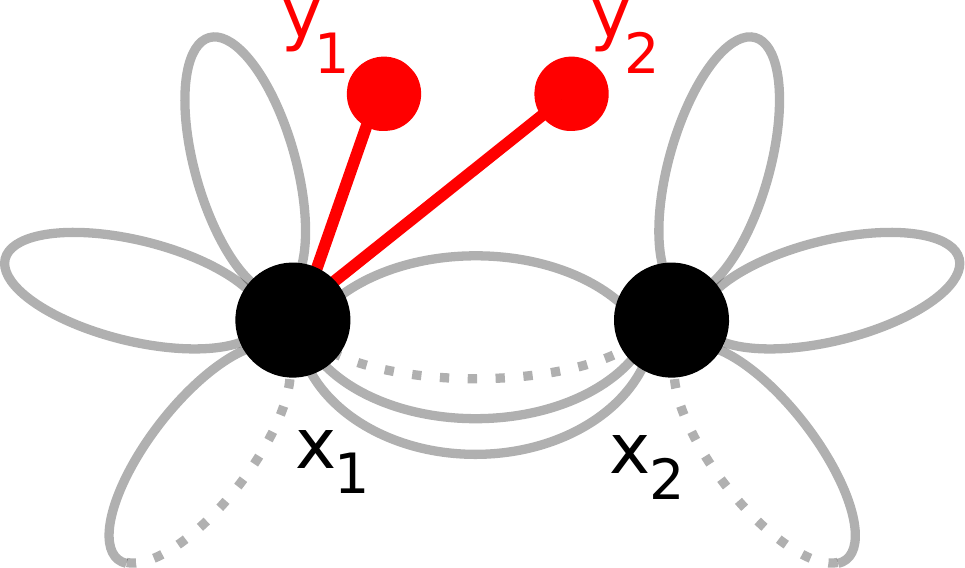}\label{F3d}}
\hfill\null
\caption{Graphical representation of terms contributing to the functional
integrals in (\ref{E48}). In \ref{F3a}, $y_1$ ($y_2$) connects to the point
$x_1$ ($x_2$), with the remaining number of points connecting $x_1$ to $x_2$.
In \ref{F3b} the connection of $y_1$ ($y_2$) is to the points at $x_2$ ($x_1$),
with the remaining points connecting to $x_1$ to $x_2$. In \ref{F3c} and
\ref{F3d} both $y_1$ and $y_2$ are connected to the same point, either $x_1$ or
$x_2$, with these being connected.  $x_1$ and $x_2$ are integrated over.} 
\end{figure*} 

In the $\vep$ expansion of $G_2(y_1,y_2)$, the coefficient proportional to
$\vep^2$, has the form $G_{2,2}=A_{2;1}+A_{2;2}$, where
\begin{eqnarray}
A_{2;1} &=& -\frac{g\mu_0^2}{4Z_0}\int\cD\phi\,e^{-\int d^Dx\,\cL_0}\phi(y_1)
\phi(y_2)\nonumber\\
&&\times\int d^Dx\,\phi^2(x)\log^2\big[i\mu_0^{1-D/2}\phi(x)\big],\label{E47}\\
A_{2;2} &=& \frac{g^2\mu_0^4}{8Z_0}\int\cD\phi\,e^{-\int d^Dx\,\cL_0}\phi(y_1)
\phi(y_2)\nonumber\\
&&\times\Big\{\int d^Dx\,\phi^2(x)\log\big[i\mu_0^{1-D/2}\phi(x)\big]\Big\}^2.
\label{E48}
\end{eqnarray}
For $A_{2;1}$ the even terms in the expansion of the imaginary logarithm give
rise to the connected terms joining $y_1$ and $y_2$, and thus the functional
integrals that arise correspond to those depicted in Fig.~\ref{F2}, where $x$
is an internal vertex that is integrated over. A detailed calculation gives
$$A_{2;1}=g\mu_0^2 K_2\int d^Dx\,\Delta(y_1-x)\Delta(x-y_2),$$
where 
\begin{widetext}
$$K_2=-\tfrac18\Big(\log^2[2\mu_0^{2-D}\Delta(0)]+2\log[2\mu_0^{2-D}\Delta(0)]
\big[\psi\big(\tfrac 32\big)+1\big]+\psi^\prime\big(\tfrac 32\big)+
\psi^2\big(\tfrac 32\big)+2\psi\big(\tfrac 32\big)-\pi^2\Big).$$

For $A_{2;2}$, the square of the integral in (\ref{E48}) leads to two internal
points $x_1$ and $x_2$ to which $y_1$ and $y_2$ can be connected. The possible
connected diagrams are shown in Fig.~\ref{F3a}-\ref{F3d}. A detailed calculation
yields
$$A_{2;2}=\frac{g^2\mu_0^2}{m^2}
\Big[K_3\int\! d^Dx\Delta(y_1-x)\Delta(x-y_2)
+\int\!\int d^Dx_1\,d^Dx_2\Delta(x_1)\Delta(x_2+y_1-y_2)f(x_1-x_2)\Big],$$
where
\begin{eqnarray}
K_3&=&\tfrac{m^2\mu_0^2\Delta(0)}4\int d^Dx\Big\{\sin^{-1}\big[\tfrac{\Delta(x)}
{\Delta(0)}\big]\big(\sin^{-1}\big[\tfrac{\Delta(x)}{\Delta(0)}\big]-\pi\big)
+\tfrac{\Delta(x)}{\Delta(0)}\sqrt{1-\big[\tfrac{\Delta(x)}{\Delta(0)}\big]^2}
\big(2\sin^{-1}\big[\tfrac{\Delta(x)}{\Delta(0)}\big]-\pi\big)\nonumber\\
&&+\big[\tfrac{\Delta(x)}{\Delta(0)}\big]^2\big(\log[2\mu_0^{2-D}\Delta(0)]
+\psi\big(\threehalf\big)-2\big)\Big\},\nonumber
\end{eqnarray}
is a dimensionless constant and
\begin{eqnarray}
f(x)&=& \half m^2\mu_0^2\Delta(x)\Big\{\sin^{-1}\big[\tfrac{\Delta(x)}
{\Delta(0)}\big]\big(\sin^{-1}\big[\tfrac{\Delta(x)}{\Delta(0)}\big]-\pi\big)
+\half\big(\log\big[2\mu_0^{2-D}\Delta(0)\big]+{\psi\big(\threehalf\big)}+1
\big)^2-2\Big\}\nonumber\\
&&+m^2\mu_0^2\Delta(0)\sin^{-1}\big[\tfrac{\Delta(x)}{\Delta(0)}\big]
\sqrt{1-\big[\tfrac{\Delta(x)}{\Delta(0)}\big]^2}+\half{m^2\mu_0^{2}\Delta(0)}
\pi\Big(1-\sqrt{1-\big[\tfrac{\Delta(x)}{\Delta(0)}\big]^2}\Big)\nonumber
\end{eqnarray}
has dimension (mass)$^D$. Collecting all terms for $G_{2,0}$,
$G_{2,1}$, and $G_{2,2}$, we obtain $G_2(y_1,y_2)=G_2(y_1-y_2)$:
\begin{equation}
G_2(y_1-y_2)=\Delta(y_1-y_2)-\vep g\mu_0^2 K_1 Q_1(y_1-y_2)+\vep^2 g\mu_0^2\Big(
(K_2+gm^{-2}K_3)Q_1(y_1-y_2)+gm^{-2}Q_2(y_1-y_2)\Big)+{\rm O}(\vep^3),
\label{E49}
\end{equation}
\end{widetext}
where we have abbreviated
$$Q_1(y_1-y_2)=\textstyle{\int} d^Dx\,\Delta(y_1-x)\Delta(x-y_2),$$
\begin{eqnarray}
Q_2(y_1-y_2)&=&\textstyle{\int\!\int}d^Dx_1\,d^Dx_2\,\Delta(x_1)\nonumber\\
&\times&\Delta(x_2+y_1-y_2)f(x_1-x_2).\nonumber
\end{eqnarray}
Since $Q_1(y_1-y_2)$ and $Q_2(y_1-y_2)$ are convolutions, their Fourier
transforms give products of the Fourier transforms of their components. Thus,
the Fourier transform of $G_2(y_1-y_2)$ takes the form
\begin{eqnarray}
\hat G_2(p) &=& \widehat\Delta(p)-\vep g\mu_0^2 K_1\widehat\Delta^2(p)
\nonumber\\
&&\!\!\!\!\!\!\!\!\!\!\!+\vep^2 g\mu_0^2[K_2+g m^{-2}K_3+gm^{-2}\hat f(p)]
\widehat\Delta^2(p).\nonumber
\end{eqnarray}

The renormalized mass $M_R^2=\lim_{p\to 0}\hat G_2^{-1}(p)$ is then
\begin{eqnarray}
M_R^2 &=& (m\mu_0)^2+\vep g\mu_0^2 K_1\nonumber\\
&&+\vep^2 g\mu_0^2[gm^{-2}(K_1^2- K_3-\hat f(0))-K_2],\nonumber
\end{eqnarray} 
which reduces to (\ref{E19}) to first order in $\vep$. On substituting $m^2=g+
\mu^2/\mu_0^2$, we get
$$ M_R^2=\mu^2+g\mu_0^2[1+\vep K_1+\vep^2(gm^{-2}(K_1^2-K_3-\hat f(0))-K_2)].$$
 
{\it Divergence structure as $D\to2$}. The divergence structure of $G_2(y_1-
y_2)$ when $D\to2$ ($\delta\to0$) can be determined from (\ref{E49}). In this
limit, $\Delta(x)\to f_1(x)+\delta f_2(x)$ while $K_1$ diverges as $-\half\log
\delta+\tfrac 32-\tfrac\gamma 2-\tfrac12 \log(4\pi)$ and $K_2$ and $\hat f(0)$
introduce divergences of order $(\log\delta)^2$. Putting this together, we find
that 
\begin{eqnarray}
G_2(y_1-y_2) &\sim& c_1+\vep(c_2+\log\delta)\nonumber\\
&&\!\!\!\!\!\! +\vep^2[c_3+\log\delta+\log^2\delta+{\rm O}(\delta)],\nonumber
\end{eqnarray}
where constant prefactors are suppressed and $c_n$ are constants. Evidently,
higher-order terms in the $\vep$ expansion for $G_2(y_1-y_2)$ display the same
algebraic divergence as the lowest-order term, with higher-order corrections
being logarithmic. This was previously observed for $G_1$ and we believe it to
be generally true.
 
In summary, if we expand $G_2(y_1-y_2)$ first in terms of $\vep$, treat $\delta$
as finite, perform a Fourier transform, and invert it to identify a renormalized
mass, we see that each coefficient in the $\vep$ expansion for $M_R^2$ also
diverges. The term in $M_R^2$ proportional to $\vep$ diverges as $\log\delta$,
while the structure of the terms proportional to $\vep^2$ introduces divergences
of up to $\log^2\delta$. Thus, in such a perturbative calculation, the mass
counterterm $\mu$ must absorb divergences that arise at each order of $\vep$.

\subsection{Three-point Green's function in second order}
\label{s2c}
The connected three-point Green's function can also be calculated up to second
order, using the techniques of the last sections. As this is tedious, we only
give final results. The first-order coefficient in the $\vep$ expansion is
\begin{equation}
G_{3,1} (y_1,y_2,y_3)=-ig\mu_0^2\sqrt{\tfrac{\pi}{2\Delta(0)}}R(y_1,y_2,y_3),
\label{E50}
\end{equation}
where 
$$R_1(y_1,y_2,y_3)=\textstyle{\int}d^D x\Delta(x-y_1)\Delta(x-y_2)\Delta(x-y_3).
$$
In the limit $D\to2$, $\delta\to 0$, the behavior of $G_{3,1}$ is determined by 
the factor $[\Delta (0)]^{-1/2} \sim \delta^{1/2}$ in (\ref{E50}), since 
$R_1 (y_1,y_2,y_3) \sim {\textrm {constant}} + O(\delta)$.

The $\vep^2$ coefficient of $G_3$ in the $\vep$ expansion is 
\begin{widetext}
\begin{eqnarray}
G_{3,2}(y_1,y_2,y_3) &=& -\tfrac{ig\mu_0^2}2 \sqrt{\tfrac{\pi}{2\Delta(0)}}
\Big\{\big(\log\big[2\mu_0^{2-D}\Delta(0)\big]+\psi(2)+2\big)R_1(y_1,y_2,y_3)
-g\mu_0^2R_2(y_1,y_2,y_3)\nonumber \\
&&-{g\mu_0^2}R_1(y_1,y_2,y_3)\Big[\Delta(0)\textstyle{\int}d^Dx
\Big(1-\big[\tfrac{\Delta(x)}{\Delta(0)}\big]^2\Big)
\log\big[1+\tfrac{\Delta(x)}{\Delta(0)}\big]\nonumber\\ 
&& \qquad+(m\mu_0)^{-2}\Big(1+\tfrac 14(2-D)\big(\log\big[2\mu_0^{2-D} 
\Delta(0)\big]+\psi\big(\tfrac 32\big)\big)\Big)\Big]\Big\},
\label{E51}
\end{eqnarray}
where
\begin{eqnarray}
R_2(y_1,y_2,y_3) &=& \textstyle{\int\int}d^D x_1\,d^Dx_2\,\big[\Delta(x_1-y_1)
\Delta(x_1-y_2)\Delta(x_2-y_3)+\Delta(x_1-y_1)\Delta(x_2-y_2)\Delta(x_1-y_3)
\nonumber \\
&&\qquad\qquad\qquad+\Delta(x_2-y_1)\Delta(x_1-y_2)\Delta(x_1-y_3)\big]
\nonumber\\
&& \times\Big\{\Delta(x_1-x_2)\big[\log\big[2\mu_0^{2-D}\Delta(0)\big]+\psi
\big(\tfrac32\big)-1\big)+2[\Delta(0)+\Delta(x_1-x_2)]\log\big[1+ 
\tfrac{\Delta(x_1-x_2)}{\Delta(0)}\big]\Big\}\nonumber
\end{eqnarray}
\end{widetext}
An analysis of  (\ref{E51}) shows that the second-order contribution to the
expansion of $G_3(y_1,y_2,y_3)$ in $\vep$ goes as $\sqrt\delta$ and introduces
corrections of order $\sqrt\delta\log\delta$. 

Thus, we see that if the first three connected Green's functions, $G_1$, $G_2
(y_1-y_2)$, and $G_3(y_1,y_2,y_3)$ are expanded as series in powers of $\vep$,
the coefficients in the series have the same {\it algebraic} behavior as $\delta
\to0$, but that the higher-order coefficients introduce additional powers of
$\log\delta$. In general, 
\begin{equation}
G_p(y_1,\ldots,y_p)\sim\delta^{p/2-1}\times[1+{\rm powers~of}~\log\delta].
\label{E52}
\end{equation}
This result is surprising: We are able to shift $G_1$ by adding a counterterm
$iv(\vep)\phi$ and we can perform mass renormalization of $G_2$ to second order
but (\ref{E52}) implies that all other higher-order Green's functions with $p
\ge3$ must vanish. As shown explicitly in Sec.~\ref{s2a} for $G_1$, this
behavior is unaffected by including the counterterm $iv(\vep)\phi$. This
suggests that the theory becomes noninteracting in the limit $D\to2$ to any
finite order in $\vep$.
 
\section{Multiple-scale analysis as $\delta\to0$}
\label{s3}
An alternative method of approaching the limit $D\to2$ ($\delta\to0$) is to
perform the sums to all orders $n$ in $\vep$ before taking the limit $\delta\to
0$. This approach is inspired by the techniques of multiple-scale perturbation
theory (MSPT), which is a powerful perturbative technique that was first used in
early calculations of planetary orbits. In conventional perturbative expansions
higher orders depend resonantly on lower orders, and as a result, the higher
orders in a perturbation expansion contain {\it secular} terms. Because secular
terms are large they tend to violate rigorous bounds that can be established
from general principles, such as conservation of energy.

A simple example is provided by the classical anharmonic oscillator, whose
Hamiltonian is $p^2+x^2+\vep x^4$. A conventional perturbative solution to the
classical equation of motion has the form $x(t)=a_0(t)+a_1(t)\vep+a_2(t)\vep^2+
...$. The coefficient $a_0(t)$ is oscillatory and thus is bounded. However,
$a_1(t)$ is secular and grows linearly with time $t$, and this growth violates
the energy conservation. The next term in the perturbation series grows
quadratically with time, and in general $a_n(t)\sim t^n$ as $t\to\infty$. It is
possible to repair this inconsistency by summing the most secular contributions
to $a_n(t)$ to all orders in powers of $\vep$, and when we do so we find that
the sum exponentiates to give a term of the form $e^{-Ct}$, where $C$ is a
constant, which no longer violates the conservation of energy \cite{BO} and
gives an accurate approximation to $x(t)$ that is valid for long times $t\gg1$.

The techniques of MSPT can also be used for quantum systems. These techniques
yield good numerical results when applied to the wave function for the quantum
anharmonic oscillator \cite{W1,W2}. The MSPT approach has also been used in
quantum field theory: It was used in perturbative QED by H.~Cheng and T.~T.~Wu
to sum over and eliminate leading-logarithm divergences that violate the
high-energy Froissart bound \cite{W3}. It was also used by L.~Dolan, R.~Jackiw,
E.~Braaten, and R.~Pisarski to sum leading infrared divergences \cite{W4}.

In this paper we apply MSPT techniques to the $\cPT$-symmetric Lagrangian
(\ref{E17}) and we show what happens if we sum to orders in $\vep$ the most
divergent perturbative contributions to the Green's functions. These
contributions are logarithmic; this is not surprising because we are expanding
in powers of $\vep$, which is a parameter in the exponent.

\begin{figure}[t]
\centering
\includegraphics[scale = 0.8]{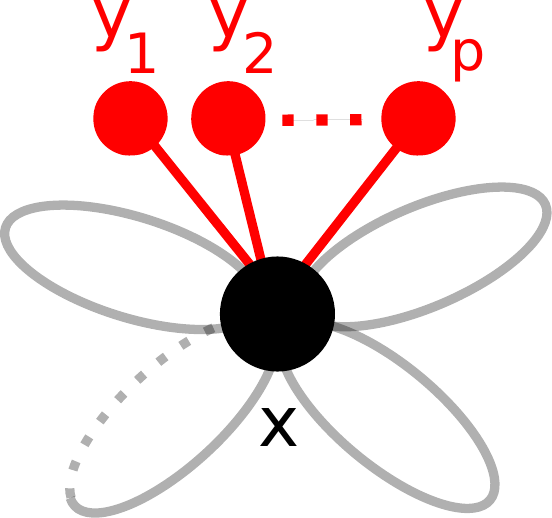}
\caption{Graphical structure of contributions to the $n$th order coefficient 
$G_{p,n}(y_1, ..., y_p)$ connected to one internal point $x$.}
\label{F4}
\end{figure}

We can identify the leading terms in the $\vep$ expansion of the Green's
functions by restricting our attention to evaluating terms for the $p$-point
connected Green's function that connect to only one internal point $x$. Then,
the expansion for the Green's function includes only those terms arising from
\begin{eqnarray}
G_{p,0}&\!=\!&\frac 1{Z_0}\int\!\cD\phi\,e^{-\int d^Dx\,\cL_0}\phi(y_1)\ldots
\phi(y_p),\nonumber\\
G_{p,n}&\!\simeq\!&\frac{g\mu_0^2}{n!Z_0}\int\!\cD\phi\,e^{-\int d^D x\,\cL_0}
\phi(y_1)\ldots\phi(y_p)I_n,\nonumber
\end{eqnarray}
with $I_n=-\half{\textstyle\int}d^Dx\,\phi^2(x)\log^n\big[i\mu_0^{1-D/2}
\phi(x)]$ as used earlier. This corresponds to diagrams of the form given in
Fig.~\ref{F4}. Converting the complex logarithm in $I_n$ to a sum of real and
imaginary terms via (\ref{E4}), employing the binomial expansion, and replacing
the real logarithm by using the replica trick, we arrive at the expression
\begin{widetext}
\begin{eqnarray}
G_{p,n}(y_1,...,y_p)&\simeq&-\frac{g\mu_0^2}{2^{n+1} n!}\lim_{N\to0}\int d^D x
\sum_{m=0}^n (i\pi)^m \binom{n}{m}\Big(\frac{d}{dN}\Big)^{n-m} \nonumber\\
&& \times\int\frac{\cD\phi}{Z_0} e^{-\int d^D x \cL_0}\phi(y_1) ...
\phi(y_p) \phi^2(x)\Big(\frac{|\phi(x)|}{\phi(x)} \Big)^m[\mu_0^{1-D/2}
\phi(x)]^{2N}\quad(n\ge1). \nonumber
\end{eqnarray}

As in Sec.~\ref{s2b} we can replace $(|\phi|/\phi)^m$ by a representation that
only contains powers of $\phi$. The resulting functional integral can be
evaluated, yielding
\begin{eqnarray}
G_{p,n}(y_1,...,y_p)&\simeq&-\frac{g\mu_0^2}{2^{n} n! }
\frac{\Delta(0)}{\sqrt\pi} \Bigl[\frac 2{\Delta(0)}\Bigr]^{p/2} 
\int d^D x \prod_{i=1}^{p}\Delta(y_i-x)\nonumber\\
&\times& \lim_{N\to0} \sum_{\substack{m=0\\m-p = {\rm even}}}^n
(i\pi)^m\binom{n}{m}\Big(\frac{d}{dN}\Big)^{n-m} [2\mu_0^{2-D}\Delta(0)]^N
\frac{\Gamma(N+\threehalf)\Gamma(N+2)}{\Gamma(N+2-p/2)}.
\label{E53}
\end{eqnarray}
\end{widetext}

Now it is possible to sum all $n$ contributions in the expansion in $\vep$ 
to form $G_p(\vep)=\sum_n G_{p,n}\vep^n$.

{\it Case $p=1$.} Evaluating (\ref{E53}) for $p=1$ we find that the coefficients
can be expressed as
\begin{eqnarray}
G_{1,n} &=& -\tfrac{igm^{-2}}{2^n n!}\big[\tfrac{2\Delta(0)}{\pi}\big]^{1/2} 
\lim_{N\to0}\big(\tfrac{d}{dN}\big)^n\nonumber\\
&&\!\!\!\! \times~\big[2\mu_0^{2-D}\Delta(0)\big]^N\sin(\pi N)\Gamma(N+2).
\label{E54}
\end{eqnarray}
Thus, $G_1(\vep)=\sum_n G_{1,n}\vep^n$ follows immediately as
\begin{eqnarray}
G_1(\vep)&=&-\tfrac{igm^{-2}}{\sqrt\pi}\big[2\mu_0^{2-D}\Delta(0)\big]^{(\vep
+1)/2}\mu_0^{-1+D/2}\nonumber\\
&&\quad\times\,\Gamma\big(2+\tfrac\vep 2\big)\sin\big(\tfrac{\pi\vep}2\big)
\nonumber
\end{eqnarray}
because $G_{1,n}$ in (\ref{E54}) are the coefficients of a Taylor expansion
about $N+\frac{\vep}{2}$. [That is, the exponential of a derivative is a
translation operator: $\exp\big(\frac{\vep}{2}\frac{d}{dN}\big)f(N)=f\big(N+
\frac{\vep}{2}\big)$.]

Now, in the limit $\delta\to0$, we get
$$G_1(\vep)\sim-\frac{igm^{-2}}{\pi^{(\vep+2)/2}}\delta^{-(\vep+1)/2}
\Gamma\big(2+\tfrac\vep 2\big)\sin\big(\tfrac{\pi\vep}2\big).$$
From this equation we see that in this multiple-scale approximation the
summation leads to an {\it algebraic} structure for the divergence as $\delta\to
0$ that becomes more pronounced with increasing $\vep$. As expected, expanding
this result for small values of $\vep$ gives the previously obtained result
$G_1(\vep)\to-igm^{-2}\delta^{-1/2}\vep/2$ plus terms containing the same power
of $\delta^{-1/2}$ multiplied by logarithms of $\delta$. 

Similarly, we can evaluate $G_2(\vep)$ in this limit. From (\ref{E53}) the
expansion coefficients become
\begin{eqnarray}
&&G_{2,n}=-\frac{g\mu_0^2}{2^{n-1} n!\sqrt\pi}\int d^Dx\,\Delta(x-y_1)
\Delta(y_2-x)\nonumber \\
&&\!\!\!\!\!\!\times~\lim_{N\to0}\big(\tfrac{d}{dN}\big)^n (N+1)\big[2\mu_0^{
2-D}\Delta(0)\big]^N\cos(\pi N)\Gamma\big(N+\threehalf\big)\nonumber
\end{eqnarray}
for $n\ge1$. The closed-form solution for the connected two-point Green's
function in this approximation is then
\begin{eqnarray}
G_2 (\vep) &=& \Delta(y_1-y_2)+gm^{-2}\big(1-\tfrac D2\big)\Delta(0)\nonumber\\
&&\!\!\!\!\!\!\!\!\!\!\!\!\!\!\!\!\!\!\times~\Big[1-\big(\tfrac{2+\vep}{\sqrt
\pi}\big)\big[2\mu_0^{2-D}\Delta(0)\big]^{\vep/2}\cos\big(\half\pi\vep\big)
\Gamma\big(\tfrac\vep2+\threehalf\big)\Big],
\nonumber
\end{eqnarray}
where we have suppressed the spatial argument $(y_1-y_2)$ of $G_2(\vep)$. In the
limit of small $\vep$, we recover (\ref{E49}) to linear order and in particular
in the double limit, taking $\delta\to0$. This reconfirms the previous results,
but also demonstrates that the two-point Green's function has an algebraic
$\delta$ dependence, given as
\begin{eqnarray}
G_2 (\vep) &=& \tfrac 1{2\pi}K_0(m\mu_0|y_1-y_2|)+\tfrac 1{4\pi}{gm^{-2}}
\nonumber\\
&&\!\!\!\!\!\!\times~\Big[1-\tfrac{2+\vep}{\sqrt\pi}(\pi\delta)^{-\vep/2}\cos
\big(\half\pi\vep\big)\Gamma\big(\tfrac\vep2+\threehalf\big)\Big]\nonumber
\end{eqnarray}
for arbitrary $\vep$. As $\vep$ increases, the divergence in $G_2(\vep)$ also
becomes more pronounced.

This procedure applies to higher-order Green's functions. The summation over $n$
with coefficients $G_{p,n}$ from (\ref{E53}) can be performed, leading to
\begin{eqnarray}
G_p(\vep) &\simeq& -g\mu_0^2\frac{\Delta(0)}{\sqrt\pi} \Big(\frac 2{\Delta(0)}
\Big)^{p/2}\int d^D x \prod_{i=1}^{p}\Delta(x-y_i)\nonumber \\
&\times&\big [2\mu_0^{2-D}\Delta(0)\big]^{\vep/2} \frac{\Gamma\big(\half\vep+
\threehalf\big)\Gamma\big(\half\vep+2\big)}{\Gamma\big(\half\vep+2-\frac p2\big)
}\nonumber\\
&\times&\left\{\begin{array}{ll} i \sin (\pi\vep/2) & p \quad {\rm odd} \\
\cos(\pi\vep/2) & p \quad{\rm even}\end{array}\right.
\nonumber
\end{eqnarray}
in which the spatial arguments of $G_p(\vep)$ have again been suppressed.
From this, it is evident that for $p>2$
$$G_p(\vep) \sim \delta^{p/2-1-\vep/2}\quad(\delta\to0).$$

\section{Summary and outlook}
\label{s4}
The Euclidean Lagrangian $\cL=\half(\nabla\phi)^2+\half\mu^2\phi^2+\half g
\mu_0^2\phi^2\big(i\mu_0^{1-D/2}\phi)^\vep$, as a field-theoretic generalization
of the quantum-mechanical Lagrangian $\cL_{qm}=\half(\nabla x)^2+\half x^2(ix
)^2$, is a laboratory for the study of $\cPT$-symmetric bosonic field theories.
We have calculated the connected Green's functions $G_1(\vep)$, $G_2(\vep;\,y_1
-y_2)$, and $G_3(\vep;\,y_1,y_2,y_3)$ to second order in an expansion in powers
of $\vep$, with an emphasis on examining the limit of this expansion as the
spacetime dimension approaches 2 from below; that is as $\delta=2-D\to0$.
We have shown that divergences appear in this limit: Specifically, to first
order in $\vep$ the expansion coefficients in the sum $G_p=\sum G_{p,n}\vep^n$
go as $G_{p,n}\sim\delta^{p/2-1}$. Thus, we observe algebraic divergences for
$p=1$. To second order in $\vep$, we find that the algebraic structure in
$\delta$ remains intact and the only changes involve logarithmic corrections.

We have attempted a perturbative renormalization scheme in which a counterterm
of the form $i v(\vep)\phi$ is introduced. This introduces a shift that makes
$G_1(\vep)$ finite but evidently this does not modify the $\delta$-dependence of
higher-order Green's functions. We have demonstrated this explicitly with our
calculation of $G_2(\vep)$. On the other hand, including an explicit mass
parameter $\mu_0$ allows for a mass renormalization, which is obtained from the
two-point Green's function $G_2$.

Taking a completely different approach, we have calculated $G_p(\vep)$ to all
orders in $\vep$ in a leading-log expansion in the context of multiple-scale 
perturbation theory. In this approximation the logarithmic divergences in
$\delta$ sum to yield an algebraic result.

Both results presented here, the perturbative calculation in $\vep$ and the
leading-logarithm sum to all powers in $\vep$, can be reconciled for small
values of $\vep$. However, it does not yet answer the question as to why in the
perturbative calculation the Green's functions of the order $p\ge3$ appear to
vanish in the limit $\delta\to0$. Of course, the work presented here does not
imply that the full theory becomes noninteracting as $D\to2$, but it is clear
that additional work is required to develop a more robust procedural approach.

\acknowledgments
CMB thanks the Alexander von Humboldt and Simons Foundations and the UK
Engineering and Physical Sciences Research Council for financial support.

\end{document}